\pdfoutput=1
\documentclass[journal]{IEEEtran}
\usepackage{stmaryrd}
\usepackage{lipsum}  
\usepackage{svg}
\makeatletter
\def\input@path{{addons/}{../addons/}}
\makeatother
\usepackage{xcolor}
\usepackage{amssymb,amsmath,amsfonts,wasysym,latexsym}
\usepackage[acronyms,shortcuts]{glossaries}
\usepackage{bm}
\usepackage{xcolor}
\usepackage{tikz}
\usetikzlibrary{patterns,arrows,decorations.pathreplacing}
\usepackage{psfrag}
\usepackage{amssymb}
\newcommand{\argmin}{\mathop{\rm argmin}}

\usepackage{epstopdf}
\usepackage{amsmath}
\usepackage{scalefnt}
\usepackage{booktabs}
\usepackage{multirow}
\usepackage{array}
\usepackage{color}
\usepackage{xcolor}
\usepackage{algorithm}
\usepackage{algpseudocode}
\usepackage{cite}
\usepackage{caption}
\usepackage{subcaption}
\usepackage{caption}
\usepackage{multirow}
\usepackage{graphicx}
\usepackage{siunitx}
\usepackage{tabularx}
\usepackage{array}
\newcolumntype{Y}{>{\raggedright\arraybackslash}X}

\definecolor{forestgreen}{rgb}{0, 0.5,0.5} 
\definecolor{darkgreen}{rgb}{0,0.392157,0} 

\newcommand{\nmode}[2]{[#1]_{(#2)}}

\newcommand{\bb}[1]{\mathbb{#1}}
\newcommand{\ten}[1]{\boldsymbol{\mathcal #1}}
\newcommand{\ma}[1]{\boldsymbol{#1}}
\definecolor{green}{rgb}{0.1,0.75,0.2}

\renewcommand{\vec}[1]{\text{vec}\big( #1\big)}
\newacronym{2G}{2G}{second generation}
\newacronym{3G}{3G}{third generation}
\newacronym{4G}{4G}{fourth generation}
\newacronym{5G}{5G}{fifth generation}
\newacronym{B5G}{B5G}{beyond fifth generation}
\newacronym{6G}{6G}{sixth generation}
\newacronym{3GPP}{3GPP}{3$\text{rd}$~Generation Partnership Project}
\newacronym{LTE}{LTE}{long term evolution}
\newacronym{NR}{NR}{new radio}
\newacronym{LS}{LS}{least squares}

\newacronym{UPA}{UPA}{uniform planar array}
\newacronym{ULA}{ULA}{uniform linear array}

\newacronym{IRS}{IRS}{intelligent reconfigurable surface}
\newacronym{RIS}{RIS}{reconfigurable intelligent surface}
\newacronym{LIS}{LIS}{large intelligent surface}
\newacronym{SDS}{SDS}{software-defined surface}

\newacronym{D2D}{D2D}{device-to-device}
\newacronym{BS}{BS}{base station}
\newacronym{UE}{UE}{user equipment}

\newacronym{SU}{SU}{single-user}
\newacronym{MU}{MU}{multi-user}
\newacronym{SISO}{SISO}{single-input single-output}
\newacronym{MISO}{MISO}{multiple-input single-output}
\newacronym{SIMO}{SIMO}{single-input multiple-output}
\newacronym{MIMO}{MIMO}{multiple-input multiple-output}

\newacronym{CSI}{CSI}{channel state information}
\newacronym{LOS}{LOS}{line-of-sight}
\newacronym{NLOS}{NLOS}{non-line-of-sight}

\newacronym{QoS}{QoS}{quality-of-service}
\newacronym{SE}{SE}{spectral efficiency}
\newacronym{EE}{EE}{energy efficiency}
\newacronym{SINR}{SINR}{signal to interference plus noise ratio}
\newacronym{SNR}{SNR}{signal to noise ratio}

\newacronym{ProSe}{ProSe}{proximity services}
\newacronym{NSPS}{NSPS}{national security and public safety}

\newacronym{RRM}{RRM}{radio resource management}
\newacronym{MS}{MS}{mode selection}
\newacronym{RA}{RA}{resource allocation}
\newacronym{PC}{PC}{power control}

\newacronym{BCD}{BCD}{block coordinate descent}

\newacronym{RF}{RF}{radio frequency}
\newacronym{AWGN}{AWGN}{additive white Gaussian noise}
\newacronym{MRC}{MRC}{maximum ratio combining}

\newacronym{AF}{AF}{amplify-and-forward}
\newacronym{DF}{DF}{decode-and-forward}
\newacronym{DFT}{DFT}{discrete Fourier transform}
\newacronym{TX}{TX}{transmitter}
\newacronym{RX}{RX}{receiver}
\newacronym{ALS}{ALS}{alternating least squares}
\newacronym{BALS}{BALS}{bilinear alternating least squares}
\newacronym{SVD}{SVD}{singular value decomposition}
\newacronym{HOSVD}{HOSVD}{high order singular value decomposition}
\newacronym{THOSVD}{THOSVD}{truncated high order singular value decomposition}
\newacronym{PARAFAC}{PARAFAC}{PARAllel FACtors}
\newacronym{AOD}{AOD}{angle of departure}
\newacronym{AOA}{AOA}{angle of arrival}
\newacronym{URA}{URA}{uniform rectangular array} 
\newacronym{ADR}{ADR}{achievable data rate}
\newacronym{NMSE}{NMSE}{normalized mean square error}
\newacronym{SER}{SER}{symbol error rate}
\newacronym{LRA}{LRA}{low-rank approximation}

\newacronym{mmWave}{mmWave}{milimiter-wave}
\newacronym{CS}{CS}{compressed sensing}
\newacronym{OFDM}{OFDM}{orthogonal frequency division multiplexing}
\newacronym{PIN}{PIN}{positive-intrinsic-negative}
\newacronym{BD-RIS}{BD-RIS}{beyond diagonal reconfigurable intelligent surface}
\newacronym{LS-Kron}{LS-Kron}{least squares Kronecker factorization}

\newacronym{BTALS}{BTALS}{block Tucker alternating least squares}
\newacronym{BTKF}{BTKF}{block Tucker Kronecker factorization}
\newacronym{PALS}{PALS}{PARAFAC alternating least squares}
\newacronym{PKF}{PKF}{PARAFAC Khatri-Rao factorization}

\newacronym{AoD}{AoD}{azimuth of departure}
\newacronym{EoD}{EoD}{elevation of departure}
\newacronym{AoA}{AoA}{azimuth of arrival}
\newacronym{EoA}{EoA}{elevation of arrival}
\newacronym{FORTE}{FORTE}{Fourth-Order Tucker Channel Estimation }
\newacronym{EORTE}{EORTE}{Eight-Order Tucker Channel Estimation}
\newacronym{CRLB}{CRLB}{Cramér–Rao Lower bound}
\newacronym{FIM}{FIM}{Fisher information matrix}
\newacronym{FORPE}{FORPE}{Fourth-Order PARAFAC Channel Estimation}
\newacronym{CDF}{CDF}{cumulative distribution function}
\newacronym{BD}{BD}{beyond-diagonal}
\newacronym{NSE}{NSE}{normalized square error}
\newacronym{TenFormer}{TenFormer}{Tensor Optimization Framework for
Beamforming, Combining, and Scattering}

\usepackage[none]{hyphenat}
\begin{document}
\title{ 
Deconstructing the Composite Channel for Beyond Diagonal RIS: Channel Estimation and Beamforming Design
 } 

\author{Fazal-E Asim, Andr\'{e}~L.~F.~de~Almeida, Bruno Sokal\\
Behrooz Makki, and Gabor Fodor}
        
 
\renewcommand{\arraystretch}{1.3} 

\maketitle


\begin{abstract}
As \ac{BD-RIS} gains increasing attention in high-frequency wireless communications, developing accurate and scalable channel-estimation methods has become both timely and essential. This paper develops a parametric channel-estimation and beamforming framework that deconstructs the composite \ac{BD-RIS} channel into its generating directional factors, thereby revealing the rich tensor structure induced jointly by propagation geometry and beyond-diagonal scattering. We propose two tensor-based estimators: \ac{FORTE}, which models the partially structured channel as a fourth-order Tucker tensor, and \ac{FORPE}, which captures the fully structured channel through a fourth-order PARAFAC model. By exploiting the partial and full channel geometry inherent to high-frequency propagation, the proposed methods deliver markedly higher estimation accuracy than state-of-the-art Least Squares and Block Tucker Kronecker Factorization approaches. In particular, \ac{FORTE} outperforms \ac{FORPE} thanks to its more compact structure, exploiting the intrinsic composite channel structure partially, given that the \ac{FORTE} method already attains an \ac{NMSE} of $10^{-4}$ at an \ac{SNR} of \SI{5}{\decibel}. While the \ac{FORPE} approach gives a unique estimation of the composite channel factor matrices, in contrast to the \ac{FORTE} method, which just gives the subspace of the composite channel factor matrices. The resulting deconstruction provides a structured representation that can be exploited not only for parametric channel estimation but also for sensing-oriented parameter extraction and tensor-structured system optimization. In addition, the \ac{TenFormer} beamforming strategy achieves spectral efficiency performance comparable to the benchmark design while significantly reducing computational complexity through parallel, tensor-structured optimization.
\end{abstract}


\renewcommand\baselinestretch{.8}

\section{Introduction}\label{Sec:Introduction}
\Ac{BD-RIS} has recently emerged as a powerful extension of conventional \ac{RIS} architectures with diagonal phase-shift matrices \cite{Jian_2022}, offering higher channel gains, improved coverage, and substantially greater control over wave propagation \cite{Hongyu_Li_2024,Li_Hongyu_2023}. By interconnecting reflecting elements through tunable components, \ac{BD-RIS} enables nonzero off-diagonal scattering matrices and, consequently, much richer electromagnetic interactions than conventional diagonal \ac{RIS}. Foundational circuit models and fully connected or group-connected architectures were introduced in \cite{Shen_Shanpu_2022}, followed by lower-complexity topologies such as forest- and tree-connected structures derived from graph-theoretic principles \cite{Nerini_Matteo_2024}. More recently, frequency-aware formulations for multiband optimization \cite{Sousa_de_Sena_2025} and hybrid transmissive, reflective, and multi-sector operating modes \cite{Hongyu_Li_2024} have further underscored the relevance of \ac{BD-RIS} for future networks.

The performance gains promised by \ac{BD-RIS} hinge on accurate \ac{CSI}, yet channel acquisition remains a major bottleneck. Because \ac{BD-RIS} operates passively, practical estimation strategies typically rely on pilot-based recovery of the composite \ac{BS}-\ac{BD-RIS}-\ac{UE} channel. In contrast to conventional \ac{RIS}, however, each \ac{BD-RIS} architecture imposes specific structural constraints on the scattering matrix while simultaneously increasing the dimensionality of the effective channel, which can lead to prohibitive training overhead \cite{Pan_Cunhua_2022}. Although the \ac{LS}-based method in \cite{Hongyu_2024} addresses group-connected and fully connected architectures, it scales poorly with the number of reflecting elements and does not exploit the block-Kronecker structure of the composite channel. Efficient and scalable \ac{CSI} acquisition for \ac{BD-RIS}-enabled systems, therefore, remains an open and pressing research problem.

Two main strategies are typically adopted to obtain instantaneous \ac{CSI} \cite{Zheng_Beixiong_2022}. The first estimates the channels between the base station (BS), \ac{UE}, and the \ac{RIS} separately by partially activating \ac{RIS} elements through dedicated RF chains. In this setting, channel reciprocity in time-division duplex systems can be exploited, and classical subspace-based techniques such as estimation of signal parameters via rotational invariance techniques, multiple signal classification \cite{Hu_Xiaoling_2022}, and compressed sensing \cite{Taha_Abdelrahman_2019} become applicable. This strategy generally incurs low training overhead regardless of the number of RIS elements and can, in principle, be extended to \ac{BD-RIS}-aided systems, since the individual channel structures do not depend on the RIS architecture. Once these channel estimates are available, existing \ac{BD-RIS} designs \cite{Li_Hongyu_2023,Nerini_Matteo_2024}, which assume perfectly known separate channels, can be directly applied.

The main drawback of the first method is the need for RF chains at the RIS, which increases hardware cost and power consumption and runs counter to the low-complexity motivation behind RIS-assisted communications. The second strategy instead estimates the cascaded or composite \ac{BS}-\ac{RIS}-\ac{UE} channel using a fully passive \ac{RIS} \cite{Swindlehurst_Lee_2022}. In this case, the \ac{BS} acquires the cascaded channel through carefully designed pilot sequences and RIS configurations, including ON/OFF-based schemes \cite{Yang_Yifei_2020} and orthogonality-based training designs \cite{Zheng_Beixiong_2020}. To further reduce training overhead, prior works have proposed anchor-assisted estimation \cite{Guan_Xinrong_2022}, which exploits the common \ac{BS}--\ac{RIS} channel shared across users, as well as sparsity- and correlation-aware methods tailored to millimeter-wave propagation \cite{Wang_Zhaorui_2020,Zhou_Gui_2022}.

Extending this passive estimation paradigm to \ac{BD}-\ac{RIS}-aided systems is considerably more challenging. The cascaded channel structure becomes tightly coupled to the underlying \ac{BD}-\ac{RIS} architecture, which makes existing methods for conventional \ac{RIS} largely inapplicable. Moreover, the inter-element coupling in \ac{BD}-\ac{RIS} substantially increases the dimension of the cascaded channel, demanding longer training sequences and new pattern-design strategies that respect architecture-dependent constraints on the scattering matrix. In addition, existing \ac{BD}-\ac{RIS} beamforming methods \cite{Li_Hongyu_Shen_2023}, which rely on separately estimated channels, cannot be directly used when only composite or cascaded channel estimates are available. These challenges strongly motivate the development of low-overhead channel-estimation methods and associated beamforming strategies tailored specifically to \ac{BD}-\ac{RIS} systems.

The work \cite{Swindlehurst_Lee_2022} investigates a variety of channel-estimation methods for \ac{RIS}-assisted communication systems under both unstructured and structured channel models. The algorithms developed for unstructured channel estimation are straightforward to implement but incur prohibitive training overhead in \ac{RIS}-based systems, thereby degrading achievable rates by requiring the estimation of a large number of channel coefficients. In contrast, geometric channel models require far fewer parameters to estimate, resulting in substantially reduced training overhead and improved estimation accuracy. These advantages, however, come at the expense of increased algorithmic complexity, the need for model-order estimation, and unavoidable modeling errors. Moving towards higher frequency bands, i.e., mmWave and THz bands, it has been shown that fewer dominant \ac{LOS} paths are present in the channel; hence, it makes more sense to use geometric channel modeling for \ac{BD}-\ac{RIS} assisted communications \cite{Abbasi_Naveed_2023}.

Tensor modeling has been successfully applied in wireless signal processing problems due to its ability to efficiently capture and exploit the inherent multidimensional structure of communication signals and channels \cite{almeida2007parafac,confac,FAVIER2012,favier2014tensor,favier2014overview,sokal_2023_beam_RIS}. More recently, tensor decompositions have also been exploited to solve the channel estimation problem in RIS-assisted communications \cite{deAraujoSAM2020,gil2021,Wei2021,Araujo_2022,nwalozie2025doubleRIS,Andre_Almeida_2025}. Specifically, \cite{gil2021} showed that the cascaded channel of RIS-assisted \ac{MIMO} systems can be arranged as a tensor, enabling decoupled estimation of the involved channels with lower training overhead. The work \cite{Araujo_2022} developed a semi-blind tensor-based formulation that jointly exploits training and the tensor signal structure, thereby improving estimation efficiency. More recently, \cite{nwalozie2025doubleRIS} extended tensor-based channel estimation to double-RIS-aided \ac{MIMO} systems by introducing coupled tensor decompositions. In the \ac{BD-RIS} context, \cite{Andre_Almeida_2025} modeled the composite channel through a block Tucker structure, showing that tensor decomposition effectively leverages the algebraic structure induced by beyond-diagonal scattering architectures to deliver estimates of the individual channels with lower training overheads compared to the LS method \cite{Hongyu_2024}. Despite these advances, all these works rely on unstructured or partially structured channel representations and do not fully exploit the parametric low-rank geometry of the underlying channel matrices.

In \cite{Asim_tvt_2025}, the multidimensional geometric structure of the \ac{BS}-\ac{RIS}-\ac{UE} channel is exploited to recast channel estimation as a single sixth-order tensor approximation problem, thereby highlighting the benefits of tensor modeling in highly structured propagation environments. The work in \cite{Asim_twc_2024} further investigates two-dimensional channel-parameter estimation in THz RIS-assisted networks by explicitly leveraging the geometric structure of the underlying channels. In particular, a structured pilot design based on the Kronecker product of horizontal and vertical training components decouples the global estimation task into lower-dimensional subproblems for each spatial direction. This reformulation converts channel estimation into a sequence of Kronecker factorization problems, which are then addressed using rank-one tensor approximation techniques, improving scalability and computational efficiency \cite{Asim_tvt_2025,Asim_twc_2024}. Despite these advances, both works rely on simplified rank-one channel representations and are therefore tailored to conventional RIS-assisted systems rather than the richer coupling structure encountered in \ac{BD-RIS} architectures.

In all the aforementioned works, the focus has been restricted either to unstructured channel models or to simplified rank-one representations tailored to conventional single-connected \ac{RIS}. As a result, the full multidimensional structure of the composite channel in \ac{BD-RIS}-assisted systems remains largely unexplored. \textcolor{black}{In contrast, to the best of our knowledge, this paper is the first to fully deconstruct the composite \ac{BD-RIS} channel into its directional factors, exposing how the \ac{BS}, \ac{UE}, and the two sides of the \ac{BD-RIS} jointly induce a rich higher-order tensor structure. This deconstruction turns the composite channel from a monolithic high-dimensional matrix into a set of interpretable propagation factors that can be directly exploited for parametric channel estimation, sensing, and system optimization.}

More specifically, rather than treating the composite channel as a generic high-dimensional object, we reveal two distinct tensorization routes, each exposing a different layer of the underlying geometry. The first route leads to Fourth-Order Tucker Channel Estimation (FORTE), in which the composite channel is cast as a fourth-order Tucker model whose modes decouple the steering matrices associated with the \ac{UE}, the \ac{BS}, and the two \ac{RIS} sides. This representation preserves the beyond-diagonal coupling while separating the main-array responses into distinct tensor modes, thereby yielding a partially structured model with a favorable balance among structural exploitation, robustness, and complexity. The second route leads to Fourth-Order PARAFAC Channel Estimation (FORPE), which goes one level deeper by explicitly unveiling the two-dimensional structure of the array manifolds. In this case, the composite channel is reformulated as a fourth-order \ac{PARAFAC} model that decouples the horizontal and vertical components of the \ac{UE}, \ac{BS}, and \ac{RIS} responses across different tensor modes, thus providing a finer parametric factorization of the propagation geometry. \textcolor{black}{Therefore, the term ``deconstructing'' is used here in a precise algebraic sense: the proposed models break the composite \ac{BD-RIS} channel into lower-dimensional directional factors whose interactions generate the observed composite channel.}

These two formulations therefore solve the same estimation problem from strikingly different perspectives: \ac{FORTE} separates the channel by terminal and surface responses, whereas \ac{FORPE} separates it by the horizontal and vertical constituents of those responses. This leads to tensor-based estimation frameworks that go beyond prior unstructured \ac{LS}-based approaches \cite{Hongyu_2024} and beyond block-structured factorization methods such as \ac{BTKF} \cite{Andre_Almeida_2025}, by incorporating a substantially richer parametric description of the composite channel itself. 

Consequently, the proposed approaches not only improve estimation accuracy in terms of \ac{NMSE}, but also establish a new modeling perspective for \ac{BD-RIS} systems in which the composite channel structure is explicitly identified, decomposed, and exploited at different levels of granularity. In this sense, the two algorithms offer different but complementary tradeoffs: \ac{FORTE} provides a more compact and computationally efficient decomposition, while \ac{FORPE} offers stronger structural exploitation, richer identifiability, and potentially higher estimation accuracy. \textcolor{black}{More broadly, this deconstructed view of the composite channel bridges channel estimation and other tasks that depend on directional information, such as localization, sensing, environment mapping, and geometry-aware beamforming.} Moreover, adopting this tensor-modeling viewpoint is useful not only for channel estimation but also for beamforming design, since it enables a unified formulation of the joint beamforming problem through which the precoders, combiners, and \ac{BD-RIS} scattering coefficients can be determined in a coordinated manner.

In summary, the contributions of the paper are as follows: 

\begin{itemize}
\item \textcolor{black}{We deconstruct the composite channel for \ac{BD}-\ac{RIS} into its generating directional factors by explicitly exploiting its multidimensional geometric structure. Unlike prior works that treat the composite channel as a single high-dimensional entity, the proposed framework separates the contributions of the \ac{BS}, \ac{UE}, and \ac{RIS} sides into physically meaningful components, thereby revealing a rich tensor structure that supports parametric channel estimation, sensing, and system optimization.}
\item Building on this decomposition, we develop two tensor-based channel-estimation frameworks:
\begin{itemize}
    \item The Fourth-Order Tucker Channel Estimation (FORTE) framework represents the composite channel as a fourth-order Tucker tensor that decouples the \ac{BS} and \ac{UE} steering matrices and the two \ac{RIS} sides into different tensor modes, yielding a compact and partially structured representation.
    \item The Fourth-Order PARAFAC Channel Estimation (FORPE) framework further exploits the two-dimensional array structure by decomposing the horizontal and vertical components of the \ac{BS}, \ac{UE}, and \ac{RIS} responses into different tensor modes. This yields a finer and more strongly structured factorization with improved identifiability properties.
\end{itemize}
\item We show that the two tensor formulations provide complementary tradeoffs between structural exploitation and complexity. In particular, FORTE offers a more compact, lower-complexity decomposition, whereas FORPE enables stronger structural identifiability through a finer tensor factorization.
\item We derive a unified tensor-based beamforming framework named Tensor Optimization Framework for Beamforming, Combining,
and Scattering (TenFormer) that jointly optimizes the BS
precoder, combiner, and BD-RIS scattering matrix using the proposed structured channel representations. By leveraging tensor-structured optimization, the proposed design achieves competitive spectral-efficiency performance while significantly reducing computational complexity.
\item We derive the \ac{CRLB} for the unstructured composite channel
as a reference benchmark and compare the proposed methods with state-of-the-art tensor \ac{BTKF}- and \ac{LS}-based estimators. Numerical results demonstrate that the proposed tensor-based approaches achieve substantial improvements in estimation accuracy and beamforming performance while maintaining similar training overhead.
\end{itemize}

\noindent \textit{\textbf{Notation}}: Matrices are represented with boldface capital letters ($\ma {A}, \ma {B}, \dots)$, and vectors are denoted by boldface lowercase letters ($\ma{a}, \ma {b}, \dots)$. Tensors are symbolized by calligraphic letters $(\ten {A}, \ten {B}, \dots)$. Transpose of a matrix $\ma {A}$ are denoted as $\ma {A}^{\textrm{T}}$. The operator $\textrm{diag}(\ma {a})$ forms a diagonal matrix out of its vector argument, while $\ast$, $\diamond$, $\otimes$ denote the conjugate, Khatri-Rao, and Kronecker products, respectively. $\ma {I}_N$ denotes the $N \times N$ identity matrix. The operator $\textrm{vec}(\cdot)$ vectorizes an $I \times J$ matrix argument, while $\textrm{unvec}_{I \times J}(\cdot)$ does the opposite operation. The tensor $\ten{Y}$ can be unfolded as three different matrices, referred to as the $1$-mode, $2$-mode, and $3$-mode unfoldings, respectively, where $[\boldsymbol{\mathcal{Y}}]_{(1)} \in \mathbb{C}^{I \times JK}$, $[\boldsymbol{\mathcal{Y}}]_{(2)} \in \mathbb{C}^{J \times IK}$, and $[\boldsymbol{\mathcal{Y}}]_{(3)} \in \mathbb{C}^{K \times IJ}$. The superscripts $\{\cdot\}^{\text{T}}$, $\{\cdot\}^{\text{*}}$, $\{\cdot\}^{\text{H}}$ and $\{\cdot\}^{\dagger }$ stand for transpose, conjugate, conjugate transpose, and pseudo-inverse operations, respectively. The operator $\Arrowvert\cdot\Arrowvert_{\text{F}}$ denotes the Frobenius norm of a matrix or tensor, and $\bb{E}\{\cdot\}$ is the expectation operator.

\begin{figure}[!t]
	\centering\includegraphics[scale=0.3]{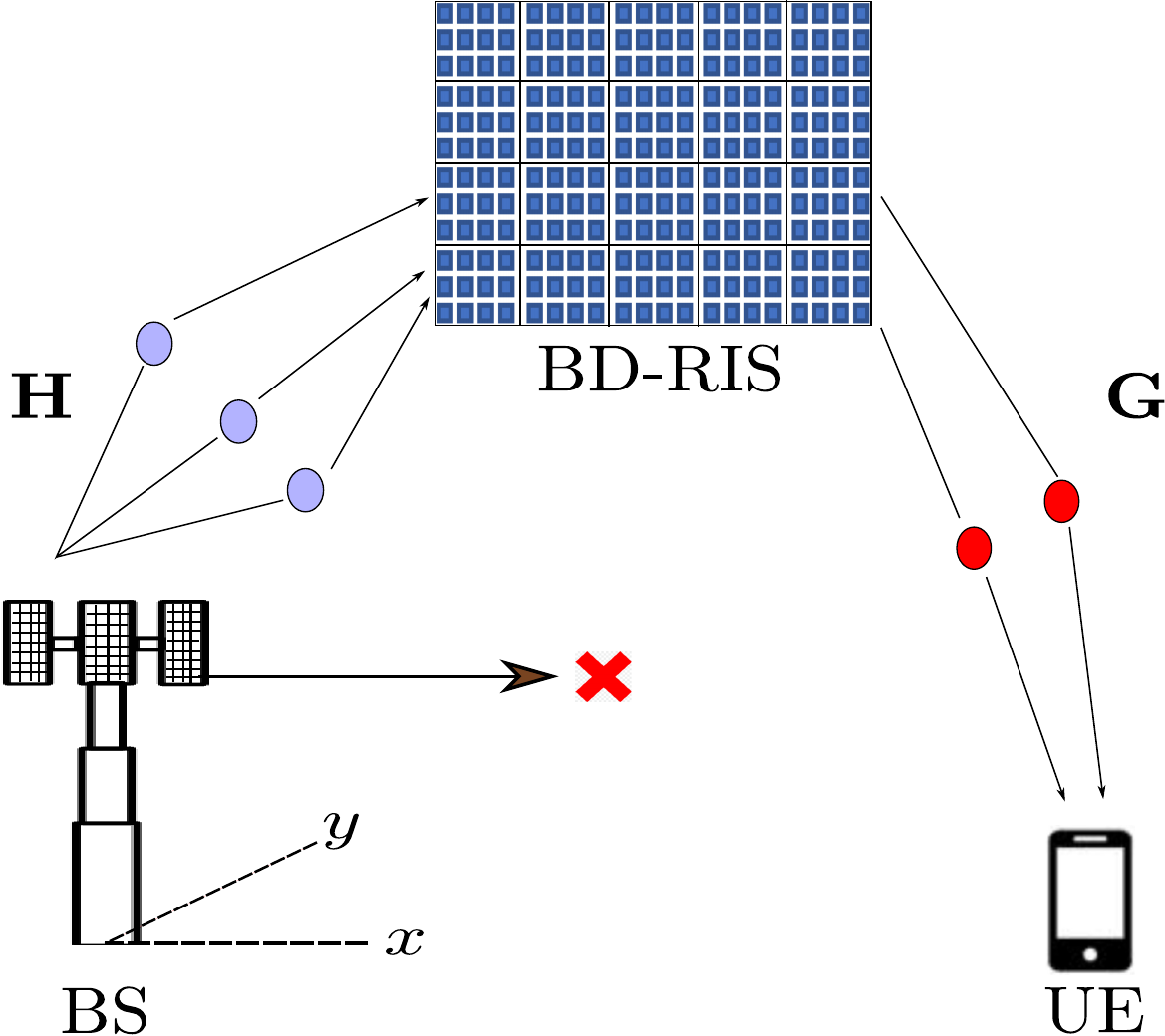}
	\caption{\small System model of a \ac{BD-RIS} assisted communications.}
	\label{fig:system_model}
\end{figure}

\section{Channel and System Model}\label{Sec:System_Model}

This section introduces the modeling framework underlying the proposed methods. We first present the structured geometric channel model for the \ac{BS}--\ac{BD-RIS} and \ac{BD-RIS}--\ac{UE} links, highlighting the parametric structure that is later exploited in the proposed tensor-based formulations. Then, the overall signal model is described, including pilot transmission, \ac{BD-RIS} training, and the resulting \ac{LS}-based composite channel estimate that serves as input to the proposed estimation algorithms.

\subsection{Structured Channel Model}
We adopt a geometric channel model for a fully connected \ac{BD-RIS}-assisted system. Owing to the limited number of dominant propagation paths at high-frequency bands, both the \ac{BS}--\ac{BD-RIS} and the \ac{BD-RIS}--\ac{UE} channels are represented using parametric angular-domain models. The channel between the \ac{BS} and the \ac{BD-RIS} is written as
\begin{align}
	\bm{H} \! = \! \sum_{r=1}^{R}\alpha_{r}\bm{b}(\phi^r_{\text{ris}_\text{A}},\theta^r_{\text{ris}_\text{A}}) \bm{a}^{\text{T}}(\phi^r_{\text{bs}},\theta^r_{\text{bs}})\nonumber\\
    \!=\! \bm{B}_{\text{RIS}} \bm{D}(\bm{\alpha}) \bm{A}_{\text{BS}}^{\text{T}} \in \mathbb{C}^{N \times M}, \label{H_sph}
\end{align}
where $\bm{a}(\phi^r_{\text{bs}},\theta^r_{\text{bs}})$ denotes the two-dimensional steering vector associated with the $r$th path at the \ac{BS}, with $\phi^r_{\text{bs}}$ and $\theta^r_{\text{bs}}$ representing its \ac{AoD} and \ac{EoD}, respectively. 
Likewise, $\bm{b}(\phi^r_{\text{ris}_\text{A}},\theta^r_{\text{ris}_\text{A}})$ denotes the corresponding steering vector at the \ac{RIS}, where $\phi^r_{\text{ris}_\text{A}}$ and $\theta^r_{\text{ris}_\text{A}}$ are the \ac{AoA} and \ac{EoA}, respectively \cite{Asim_Antreich_2021}. The coefficient $\alpha_r$ is the complex gain of the $r$th propagation path. In matrix form, $\bm{B}_{\text{RIS}} \in \mathbb{C}^{N \times R}$ collects the RIS-side steering vectors, $\bm{A}_{\text{BS}} \in \mathbb{C}^{M \times R}$ collects the BS-side steering vectors, and $\bm{D}(\bm{\alpha}) \in \mathbb{C}^{R \times R}$ is a diagonal matrix containing the path gains. Similarly, the channel between the \ac{BD-RIS} and the \ac{UE} is modeled as
\begin{align}
    \bm{G}  \!=\!\sum_{l=1}^{L}\beta_{l}\bm{q}(\phi^l_{\text{ue}},\theta^l_{\text{ue}}) \bm{p}^{\text{T}}(\phi^l_{\text{ris}_\text{D}},\theta^l_{\text{ris}_\text{D}}) \nonumber\\
    \!=\! \bm{A}_{\text{UE}} \bm{D}(\bm{\beta}) \bm{C}_{\text{RIS}}^{\text{T}} \in \mathbb{C}^{Q \times N}, \label{G_sph}
\end{align}
where $\bm{q}(\phi^l_{\text{ue}},\theta^l_{\text{ue}})$ is the two-dimensional steering vector associated with the $l$th path at the \ac{UE}, and $\phi^l_{\text{ue}}$ and $\theta^l_{\text{ue}}$ denote its \ac{AoD} and \ac{EoD}, respectively. Likewise, $\bm{p}(\phi^l_{\text{ris}_\text{D}},\theta^l_{\text{ris}_\text{D}})$ is the steering vector corresponding to the departure side of the \ac{RIS}, with $\phi^l_{\text{ris}_\text{D}}$ and $\theta^l_{\text{ris}_\text{D}}$ denoting the associated \ac{AoA} and \ac{EoA}. The coefficient $\beta_l$ is the complex gain of the $l$th path. Moreover, $\bm{C}_{\text{RIS}} \in \mathbb{C}^{N \times L}$ is the RIS steering matrix, $\bm{A}_{\text{UE}} \in \mathbb{C}^{Q \times L}$ is the receive steering matrix at the \ac{UE}, and $\bm{D}(\bm{\beta}) \in \mathbb{C}^{L \times L}$ is the diagonal matrix containing the corresponding path gains.
\textcolor{black}{
\subsection{System Model}
We consider a BD-RIS assisted MIMO communication system in which a \ac{BS} equipped with a \ac{UPA} of $M$ transmit antennas communicates with a \ac{UE} equipped with a \ac{UPA} of $Q$ receive antennas through a fully connected \ac{BD-RIS} with $N$ reflecting elements, as illustrated in Figure \ref{fig:system_model}. During training, the \ac{BD-RIS} sequentially applies $K$ known scattering configurations. For the $k$th configuration, the received signal is given by
\begin{align}
    \bm{Y}_k^\prime =  \bm{G} \bm{\Omega}^\prime_{k} \bm{H} \bm{X} + \bm{V}_k^\prime \in \mathbb{C}^{Q \times T}, \label{main_eq_k}
\end{align}
where $\bm{X}\in \bb{C}^{M \times T}$ is the pilot matrix, $\bm{H} \in \bb{C}^{N \times M}$ denotes the channel from the \ac{BS} to the \ac{BD-RIS}, $\bm{G} \in \bb{C}^{Q \times N}$ denotes the channel from the \ac{BD-RIS} to the \ac{UE}, and $\bm{\Omega}^\prime_{k} \in \bb{C}^{N \times N}$ is the known scattering matrix applied by the \ac{BD-RIS} in the $k$th training slot. The noise term $\bm{V}_k^\prime \sim \mathcal{CN}\left(\bm{0}_{Q \times T},\sigma_n^2\bm{I}_{Q \times T}\right)$ is a circularly symmetric additive white Gaussian noise matrix with variance $\sigma_n^2$. By filtering the received signal in \eqref{main_eq_k} using the known orthogonal pilot matrix, we obtain
\begin{align}
    \bm{Y}_k = \bm{Y}_k^\prime \bm{X}^{\text{H}} = \bm{G} \bm{\Omega}^\prime_{k} \bm{H} + \bm{V}_k \in \mathbb{C}^{Q \times M}, \label{filtered_signal}
\end{align}
where $\bm{X}\bm{X}^{\text{H}}=\bm{I}_N$, while $\bm{V}_k= \bm{V}_k^\prime\bm{X}^{\text{H}}$. Applying the $\vec{\cdot}$ operator to \eqref{filtered_signal} and stacking the resulting observations for all $K$ scattering configurations yield $\bm{Y}= \left[ \vec{\bm{Y}_1}, \dots, \vec{\bm{Y}_K}\right]= \left( \bm{H}^{\text{T}} \otimes \bm{G}\right) \bm{\Omega} + \bm{V} \in \mathbb{C}^{QM \times K}$,
where $\bm{\Omega} = \left[ \vec{\bm{\Omega}^\prime_{1}},\dots,\vec{\bm{\Omega}^\prime_{K}} \right] \in \bb{C}^{N^2 \times K}$ collects the vectorized \ac{BD-RIS} training matrices, and $\bm{V} = \left[ \vec{\bm{V}_{1}},\dots,\vec{\bm{V}_{K}} \right] \in \bb{C}^{QM \times K}$ is the corresponding noise matrix. 
An estimate of the composite channel $\ma{E}$ can then be obtained by right-filtering with the known \ac{BD-RIS} training matrix, coinciding with the solution to the \ac{LS} problem
\begin{align}
  \label{eq:ls_problem}  \hat{\ma{E}} = \underset{\ma{E}}{\argmin} \big\| \ma{Y} - \ma{E}\ma{\Omega}^{\text{H}}\big\|_\mathrm{F}^2,
\end{align}
where $\hat{\ma{E}} \approx \hat{\ma{H}} \otimes \hat{\ma{G}}$ denotes an estimate of the combined, or composite, channel. Hence,
\begin{align}
    \hat{\bm{E}} = \bm{Y}\bm{\Omega}^{\text{H}} = \bm{H}^{\text{T}} \otimes \bm{G} + \bm{N} \in \mathbb{C}^{QM \times N^2}, \label{E_est}
\end{align}
where $\bm{\Omega}\bm{\Omega}^{\text{H}} = \bm{I}_{N^2}$ and $\bm{N}=\bm{V}\bm{\Omega}^{\text{H}}$. 
}
In the next section, we move beyond this unstructured estimate by explicitly exploiting the intrinsic geometric structure of the composite channel, thereby enabling improved channel estimation and the design of precoding, combining, and scattering matrices.

\section{Structured Composite Channel Estimation} 

Our goal is to exploit the intrinsic geometry of the composite channel via two distinct tensorization routes, each yielding a different estimation algorithm. In the first route, followed by \ac{FORTE}, the composite channel is represented as a fourth-order Tucker model that decouples the steering matrices of the \ac{BS}, the \ac{UE}, and the \ac{RIS} into different tensor modes. In the second route, followed by \ac{FORPE}, we further expose the two-dimensional structure of the array responses and represent the same composite channel as a fourth-order \ac{PARAFAC} model that decouples the horizontal and vertical components of the \ac{BS}, \ac{UE}, and \ac{RIS} into different modes. Thus, \ac{FORTE} separates the problem at the level of the main channel factors, whereas \ac{FORPE} separates it at the finer level of the underlying horizontal and vertical array components. The Tucker and \ac{PARAFAC} formulations provide complementary tradeoffs: the Tucker model yields a more compact representation and lower-complexity decomposition, whereas the \ac{PARAFAC} model offers stronger uniqueness and identifiability properties by further decoupling the \ac{BD}-\ac{RIS} responses into horizontal and vertical components. In the following subsections, these two formulations are developed and used to derive the corresponding algorithms for estimating the factor matrices and extracting the channel parameters.

\subsection{Fourth-Order Tucker Channel Estimation (FORTE)}
The rationale behind \ac{FORTE} is to exploit a representation that preserves the physical factors of the composite channel while keeping the model sufficiently compact for robust estimation. To this end, we retain the steering matrices associated with the \ac{BS}, the \ac{UE}, and the two \ac{RIS} sides as separate building blocks. This intermediate level of factorization yields a fourth-order Tucker model, in which the channel factors are decoupled across different tensor modes. 

By substituting the geometric channel models in \eqref{H_sph} and \eqref{G_sph} into \eqref{E_est}, the composite channel can be written as follows: 
\begin{align}
    \bm{\hat{E}} = \bm{H}^\text{T} \otimes \bm{G} = \left\{ \bm{B}_{\text{RIS}} \bm{D}(\bm{\alpha}) \bm{A}_{\text{BS}}^\text{T} \right\}^\text{T}  \otimes  \left\{ \bm{A}_{\text{UE}} \bm{D}(\bm{\beta}) \bm{C}_{\text{RIS}}^\text{T} \right\}.\nonumber
\end{align} 
Rearranging the factors in the previous equation yields
  \begin{align}
     \bm{\hat{E}} = \bm{H}^\text{T} \otimes \bm{G} = \left\{ \bm{A}_{\text{BS}} \bm{D}(\bm{\alpha}) \bm{B}_{\text{RIS}}^\text{T} \right\}  \otimes  \left\{ \bm{A}_{\text{UE}} \bm{D}(\bm{\beta}) \bm{C}_{\text{RIS}}^\text{T} \right\}.\nonumber
 \end{align}
 Using identity $\left(\bm{A} \otimes \bm{B}\right) \left(\bm{C} \otimes \bm{D}\right) = \left(\bm{AC} \otimes \bm{BD}\right)$ leads to
 \begin{align}
     \bm{\hat{E}} = \left( \bm{A}_{\text{BS}} \otimes \bm{A}_{\text{UE}}\right) \left( \bm{D}(\bm{\alpha}) \otimes \bm{D}(\bm{\beta})\right) \left( \bm{B}_{\text{RIS}} \otimes \bm{C}_{\text{RIS}}\right)^\text{T} \label{tucker_model}
 \end{align}
 where
 \vspace{-2ex}
 \begin{equation}
     \bm{D}_{\bm{\alpha\beta}}
= \bm{D}(\bm{\alpha}) \otimes \bm{D}(\bm{\beta})
\in \mathbb{C}^{LR \times LR} \label{D_Core}.
 \end{equation}
 Equation \eqref{tucker_model} can be equivalently expressed as the following fourth-order Tucker tensor model:
 \begin{align}
     \boldsymbol{\hat{\mathcal{E}}}= \boldsymbol{\mathcal{D}}_{\bm{\alpha\beta}} \times_1 \bm{A}_{\text{UE}} \times_2 \bm{A}_{\text{BS}} \times_3 \bm{C}_{\text{RIS}} \times_4 \bm{B}_{\text{RIS}}, \label{Tuker_4}
 \end{align}
 where $\boldsymbol{\mathcal{E}} \in \mathbb{C}^{Q \times M \times N \times N}$ denotes the resulting fourth-order composite channel tensor, while $\boldsymbol{\mathcal{D}}_{\bm{\alpha\beta}}$ is the corresponding core tensor. The latter is obtained from the mapping in \eqref{D_Core} and contains the joint complex path gains associated with the Tucker representation. Once the model in \eqref{Tuker_4} is established, the unknown factor matrices can be estimated iteratively by means of an \ac{ALS} procedure. We begin with the \ac{UE}-side factor matrix $\hat{\bm{A}}_{\text{UE}}$, which is obtained by solving the following optimization problem:
\begin{align} \label{mode1_forte}
      \min_{\boldsymbol{\hat{A}_{\text{UE}}}} \left \| [\boldsymbol{\hat{\mathcal{E}}}]_{(1)}  -  \bm{A}_{\text{UE}} [\boldsymbol{\mathcal{D}}_{\bm{\alpha\beta}}]_{(1)} \left[\bm{B}_{\text{RIS}} \otimes  \bm{C}_{\text{RIS}} \otimes \bm{A}_{\text{BS}}\right]^\text{T} \right\|  _\text{F}^2,
\end{align}
 whose first mode unfolding is given as
 \begin{align}
     [\boldsymbol{\hat{\mathcal{E}}}]_{(1)} = \bm{A}_{\text{UE}} [\boldsymbol{\mathcal{D}}_{\bm{\alpha\beta}}]_{(1)} \left[\bm{B}_{\text{RIS}} \otimes  \bm{C}_{\text{RIS}} \otimes \bm{A}_{\text{BS}}\right]^\text{T} \mathbb{C}^{Q \times N^2M}.\nonumber
 \end{align}
 The channel factor at the \ac{BS} $\hat{\bm{A}}_{\text{BS}}$ is estimated \textit{via} the following optimization problem as:
 \begin{align} \label{mode2_forte}
      \min_{\boldsymbol{\hat{A}_{\text{BS}}}} \left \| [\boldsymbol{\hat{\mathcal{E}}}]_{(2)} -  \bm{A}_{\text{BS}} [\boldsymbol{\mathcal{D}}_{\bm{\alpha\beta}}]_{(2)} \left[\bm{B}_{\text{RIS}} \otimes  \bm{C}_{\text{RIS}} \otimes \bm{A}_{\text{UE}}\right]^\text{T} \right\|  _\text{F}^2,
\end{align}
where $\boldsymbol{[\mathcal{E}]}_{(2)}$ is the second mode unfolding of \eqref{Tuker_4}, given as
 \begin{align}
     [\boldsymbol{\hat{\mathcal{E}}}]_{(2)} = \bm{A}_{\text{BS}} [\boldsymbol{\mathcal{D}}_{\bm{\alpha\beta}}]_{(2)} \left[\bm{B}_{\text{RIS}} \otimes  \bm{C}_{\text{RIS}} \otimes \bm{A}_{\text{UE}}\right]^\text{T} \mathbb{C}^{M \times N^2Q}.\nonumber
 \end{align}
 The channel factor $\hat{\bm{C}}_{\text{RIS}}$ colecting the responses of arriving paths at the \ac{BD}-\ac{RIS} is estimated by solving: 
 \begin{align} \label{mode3_forte}
      \min_{\boldsymbol{\hat{C}_{\text{RIS}}}} \left \| [\boldsymbol{\hat{\mathcal{E}}}]_{(3)}  -  \bm{C}_{\text{RIS}} [\boldsymbol{\mathcal{D}}_{\bm{\alpha\beta}}]_{(3)} \left[\bm{B}_{\text{RIS}} \otimes  \bm{A}_{\text{BS}} \otimes \bm{A}_{\text{UE}}\right]^\text{T} \right\|  _\text{F}^2,
\end{align}
where $\boldsymbol{[\mathcal{E}]}_{(3)}$ is the third mode unfolding of \eqref{Tuker_4} given as 
 \begin{align}
     [\boldsymbol{\hat{\mathcal{E}}}]_{(3)} = \bm{C}_{\text{RIS}} [\boldsymbol{\mathcal{D}}_{\bm{\alpha\beta}}]_{(3)} \left[\bm{B}_{\text{RIS}} \otimes  \bm{A}_{\text{BS}} \otimes \bm{A}_{\text{UE}}\right]^\text{T} \mathbb{C}^{N \times MNQ}\nonumber
 \end{align}
The channel factor $\hat{\bm{B}}_{\text{RIS}}$ collecting the departuring paths from the \ac{BD}-\ac{RIS} is estimated \textit{via} the following problem: 
 \begin{align} \label{mode4_forte}
      \min_{\boldsymbol{\hat{B}_{\text{RIS}}}} \left \| [\boldsymbol{\hat{\mathcal{E}}}]_{(4)} -  \bm{B}_{\text{RIS}} [\boldsymbol{\mathcal{D}}_{\bm{\alpha\beta}}]_{(4)} \left[\bm{C}_{\text{RIS}} \otimes  \bm{A}_{\text{BS}} \otimes \bm{A}_{\text{UE}}\right]^\text{T} \right\|  _\text{F}^2,
\end{align}
where $\boldsymbol{[\mathcal{E}]}_{(4)}$ is the fourth mode unfolding of \eqref{Tuker_4} given as 
 \begin{align}
     [\boldsymbol{\hat{\mathcal{E}}}]_{(4)} = \bm{B}_{\text{RIS}} [\boldsymbol{\mathcal{D}}_{\bm{\alpha\beta}}]_{(4)} \left[\bm{C}_{\text{RIS}} \otimes  \bm{A}_{\text{BS}} \otimes \bm{A}_{\text{UE}}\right]^\text{T} \mathbb{C}^{N \times MNQ}.\nonumber
 \end{align}
 Finally, the vectorized complex-path gain can be estimated as
 \begin{align}
     \hat{\bm{d}}_{\bm{\alpha\beta}} = \left[\left( \bm{B}_{\text{RIS}} \otimes \bm{C}_{\text{RIS}}\right) \diamond \left( \bm{A}_{\text{BS}} \otimes \bm{A}_{\text{UE}}\right)\right]^\dagger \vec{{\hat{\bm{E}}}} \in \mathbb{C}^{LR \times 1}.\nonumber
 \end{align}
 where $\bm{d}_{\bm{\alpha\beta}}$ is the diagonal vector of $\bm{D}_{\bm{\alpha\beta}}$.
 The detail steps are shown in Algorithm \ref{algorithm_forte}.
 
 \begin{algorithm}[!t]
	\begingroup
	\setlength{\itemsep}{0.2ex}
	\setlength{\parsep}{0pt}
	\setlength{\topsep}{0pt}
	\setlength{\partopsep}{0pt}
	\setlength{\abovedisplayskip}{0.4ex}
	\setlength{\belowdisplayskip}{0.4ex}
	\begin{algorithmic}[1]
		\caption{Fourth-Order Tucker Channel Estimation (FORTE) Method}\label{algorithm_forte}
		\State \textbf{Inputs}: Estimated composite channel \eqref{E_est}  $\bm{\hat{E}}$, and transform it as tensor $\ten{E}$ \eqref{Tuker_4}, factorize \textit{via} \ac{ALS}:
		\State Set $i=0$. Randomly initialize $\bm{A}_{\text{BS}}$, $\bm{B}_{\text{RIS}}$, $\bm{C}_{\text{RIS}}$. 
\For{$i = 1:I$}		
		\State Compute an LS estimate of $\bm{A}_{\text{UE}}$ as 
  \begin{equation*}
      \hat{\bm{A}}_{\text{UE}} = [\boldsymbol{\hat{\mathcal{E}}}]_{(1)}\left[[\boldsymbol{\mathcal{D}}_{\bm{\alpha\beta}}]_{(1)}\left[\bm{B}_{\text{RIS}} \otimes  \bm{C}_{\text{RIS}} \otimes \bm{A}_{\text{BS}}\right]^\text{T}\right]^{\dagger}
  \end{equation*}
  	\State Compute an LS estimate of $\bm{A}_{\text{BS}}$ as
   \begin{equation*}
   \hat{\bm{A}}_{\text{BS}} = [\boldsymbol{\hat{\mathcal{E}}}]_{(2)}\left[[\boldsymbol{\mathcal{D}}_{\bm{\alpha\beta}}]_{(2)}\left[\bm{B}_{\text{RIS}} \otimes  \bm{C}_{\text{RIS}} \otimes \bm{A}_{\text{UE}}\right]^\text{T}\right]^{\dagger}
   \end{equation*}
   \State Compute an LS estimate of $\bm{C}_{\text{RIS}}$ as
   \begin{equation*}
   \hat{\bm{C}}_{\text{RIS}} = [\boldsymbol{\hat{\mathcal{E}}}]_{(3)}\left[[\boldsymbol{\mathcal{D}}_{\bm{\alpha\beta}}]_{(3)}\left[\bm{B}_{\text{RIS}} \otimes  \bm{A}_{\text{BS}} \otimes \bm{A}_{\text{UE}}\right]^\text{T}\right]^{\dagger}
   \end{equation*}
   \State Compute an LS estimate of $\bm{B}_{\text{RIS}}$ as
   \begin{equation*}
   \hat{\bm{B}}_{\text{RIS}} = [\boldsymbol{\hat{\mathcal{E}}}]_{(4)}\left[[\boldsymbol{\mathcal{D}}_{\bm{\alpha\beta}}]_{(4)}\left[\bm{C}_{\text{RIS}} \otimes  \bm{A}_{\text{BS}} \otimes \bm{A}_{\text{UE}}\right]^\text{T}\right]^{\dagger}
   \end{equation*}
   \State Compute an LS estimate of $\hat{\bm{d}}_{\alpha\beta}$ as
   \begin{equation*}
     \hat{\bm{d}}_{\alpha\beta} = \left[\left( \hat{\bm{B}}_{\text{RIS}} \otimes \hat{\bm{C}}_{\text{RIS}}\right) \diamond \left( \hat{\bm{A}}_{\text{BS}} \otimes \hat{\bm{A}}_{\text{UE}}\right)\right]^\dagger \textrm{vec}\left\{{\hat{\bm{E}}}\right\} 
   \end{equation*}
   \State 
   Compute $\hat{\ma{E}}_{(i)}= \hat{\bm{C}}_{\text{RIS}}[\hat{\boldsymbol{\mathcal{D}}}_{\bm{\alpha\beta}}]_{(3)} \left[ \hat{\bm{B}}_{\text{RIS}} \otimes  \hat{\bm{A}}_{\text{BS}} \otimes \hat{\bm{A}}_{\text{UE}} \right]^{\text{T}}$	

   and calculate the error $\epsilon_{(i)} = \big\|[\boldsymbol{\hat{\mathcal{E}}}]_{(3)}-\hat{\ma{E}}_{(i)} \big\|_{\text{F}}^{2}$,
 
    \State Check convergence and stop if $|\epsilon_{(i)} - \epsilon_{(i-1)}| \leq \eta$.
 \EndFor
		\State Return $\hat{\bm{A}}_{\text{UE}} $, $\hat{\bm{A}}_{\text{BS}}$, $\hat{\bm{C}}_{\text{RIS}} $, $\hat{\bm{B}}_{\text{RIS}}$, and $\hat{\bm{d}}_{\bm{\alpha\beta}}$.
	\end{algorithmic}
	\endgroup
\end{algorithm}
\subsection{Fourth-Order PARAFAC Channel Estimation (FORPE)}

The improved identifiability of the \ac{PARAFAC} formulation follows from standard uniqueness properties of low-rank tensor decompositions under mild rank conditions. The rationale behind \ac{FORPE} is to further refine the modeling route adopted by \ac{FORTE}. While \ac{FORTE} preserves the main channel factors associated with the \ac{BS}, the \ac{UE}, and the two \ac{RIS} sides as separate tensor modes, \ac{FORPE} goes one step deeper and explicitly unveils the two-dimensional structure of those array responses. More precisely, each steering matrix is further decomposed into horizontal and vertical components, yielding a more detailed factorization of the composite channel geometry. This finer decomposition transforms the partially structured Tucker representation in \eqref{tucker_model} into a fourth-order \ac{PARAFAC} model. With this full geometric decomposition, the composite channel can be written as follows:
\begin{align} \label{E-full_channel}
          \hat{\bm{E}} =  \left[ \left (\bm{A}^y_{\text{BS}} \diamond \bm{A}^z_{\text{BS}}\right) \otimes \left (\bm{A}^y_{\text{UE}} \diamond \bm{A}^z_{\text{UE}} \right ) \right]  \bm{D}_{\bm{\alpha\beta}}  \nonumber \\  \left[ \left(\bm{B}^y_{\text{RIS}} \diamond \bm{B}^z_{\text{RIS}}\right) \otimes \left(\bm{C}^y_{\text{RIS}} \diamond \bm{C}^z_{\text{RIS}} \right)\right]^\text{T} 
 \end{align}
 where $\bm{A}^y_{\text{BS}} \in \mathbb{C}^{M_y \times R} $, $\bm{A}^y_{\text{UE}} \in \mathbb{C}^{Q_y \times L}$, $\bm{B}^y_{\text{RIS}} \in \mathbb{C}^{N_y \times R}$, $\bm{C}^y_{\text{RIS}} \in \mathbb{C}^{N_y \times L} $,  and $\bm{A}^z_{\text{BS}} \in \mathbb{C}^{M_z \times R} $, $\bm{A}^z_{\text{UE}} \in \mathbb{C}^{Q_z \times L}$, $\bm{B}^z_{\text{RIS}} \in \mathbb{C}^{N_z \times R}$, $\bm{C}^z_{\text{RIS}} \in \mathbb{C}^{N_z \times L} $ are the horizontal ($y$) and vertical ($z$) steering matrices, respectively. Applying permutation matrices $\bm{P}_1 \in \mathbb{R}^{Q_zM_zQ_yM_y \times Q_zQ_yM_zM_y}$ and $\bm{P}_2 \in \mathbb{R}^{N_zN_zN_yN_y \times N_zN_yN_zN_y} $ to equation \eqref{E-full_channel} leads to:
 \begin{align}
          \hat{\bar{\bm{E}}} = \bm{P}_1 \left[ \left (\bm{A}^y_{\text{BS}} \diamond \bm{A}^z_{\text{BS}}\right) \otimes \left (\bm{A}^y_{\text{UE}} \diamond \bm{A}^z_{\text{UE}} \right ) \right]  \bm{D}_{\bm{\alpha\beta}} \nonumber \\ \left[ \left(\bm{B}^y_{\text{RIS}} \diamond \bm{B}^z_{\text{RIS}}\right) \otimes \left(\bm{C}^y_{\text{RIS}} \diamond \bm{C}^z_{\text{RIS}} \right)\right]^\text{T} \bm{P}_2^\text{T}
 \end{align}
 which can be further rewritten in its new formulation as: 
 \begin{align}
     \hat{\bar{\bm{E}}} =  \left [ \left( \bm{A}^y_{\text{BS}} \otimes \bm{A}^y_{\text{UE}}  \right) \diamond  \left( \bm{A}^z_{\text{BS}} \otimes \bm{A}^z_{\text{UE}}  \right) \right]  \bm{D}_{\bm{\alpha\beta}} \nonumber \\ \left[ \left( \bm{B}^y_{\text{RIS}} \otimes \bm{C}^y_{\text{RIS}} \right) \diamond \left( \bm{B}^z_{\text{RIS}} \otimes \bm{C}^z_{\text{RIS}} \right) \right]^\text{T} 
     \label{Tucker_8}
 \end{align}
 Rewriting \eqref{Tucker_8} in compact form leads to the following double-sided Khatri-Rao model:
 \begin{align}
     \hspace{-3ex} \hat{\bar{\bm{E}}} = \left[ \bm{A}_y\diamond \bm{A}_z\right] \bm{D}_{\bm{\alpha\beta}} \left[ \bm{B}_y\diamond\bm{B}_z\right]^\text{T} \in \mathbb{C}^{Q_yM_yQ_zM_z \times N_y^2N_z^2} \label{PARAFAC_4}
 \end{align}
 where $ \bm{A}_y = \bm{A}^y_{\text{BS}} \otimes \bm{A}^y_{\text{UE}} \in \mathbb{C}^{Q_yM_y \times LR}  $,  $\bm{A}_z =\bm{A}^z_{\text{BS}} \otimes \bm{A}^z_{\text{UE}} \in \mathbb{C}^{Q_zM_z \times LR}$, $\bm{B}_y = \bm{B}^y_{\text{RIS}} \otimes \bm{C}^y_{\text{RIS}} \in \mathbb{C}^{N_y^2 \times LR}$, and $\bm{B}_z=\bm{B}^z_{\text{RIS}} \otimes \bm{C}^z_{\text{RIS}}\in \mathbb{C}^{N_z^2 \times LR}$.
 Equation \eqref{PARAFAC_4} can be equivalently expressed as the following fourth-order \ac{PARAFAC} tensor model, with $\bm{\mathcal{\hat{\bar{E}}}} \in \mathbb{C}^{Q_zM_z\times Q_yM_y \times N_z^2 \times N_y^2}$:
 \begin{align} 
	\bm{\mathcal{\hat{\bar{E}}}} = \bm{\mathcal{D}}^{\bm{\alpha,\beta}}_{4,LR }\times_1 \bm{A}_z \times_2 \bm{A}_y \times_3 \bm{B}_z \times_4 \bm{B}_y. \label{FORPE_ten}
\end{align}
where $\bm{\mathcal{D}}^{\bm{\alpha,\beta}}_{4,LR } = (\bm{I}_{LR} \diamond \bm{I}_{LR}) \bm{D}_{\bm{\alpha\beta}}(\bm{I}_{LR} \diamond \bm{I}_{LR})^\text{T} \in \mathbb{C}^{LR\times LR \times LR \times LR}$ denotes the core tensor of the fourth-order \ac{PARAFAC} model. This tensor is constructed from \eqref{D_Core} and can be interpreted as an identity core whose nonzero entries are weighted by the combined complex path gains. 

Based on the tensor model in \eqref{FORPE_ten}, the unknown factor matrices are estimated iteratively by the proposed \ac{FORPE} algorithm. We start with the factor matrix $\hat{\bm{A}}_{\text{z}}$, which is obtained by solving the following \ac{LS} optimization problem:
\begingroup
\setlength{\abovedisplayskip}{0.7ex}
\setlength{\belowdisplayskip}{0.7ex}
\setlength{\abovedisplayshortskip}{0.7ex}
\setlength{\belowdisplayshortskip}{0.7ex}
\begin{align}
      \min_{\boldsymbol{\hat{A}_{\text{z}}}} \left \| [\bm{\mathcal{\hat{\bar{E}}}}]_{(1)}  -  \bm{A}_{\text{z}} \bm{D}_{\bm{\alpha\beta}} \left[\bm{B}_{\text{y}} \diamond  \bm{B}_{\text{z}} \diamond \bm{A}_{\text{y}}\right]^\text{T} \right\|  _\text{F}^2, \label{A_z}
\end{align}
\endgroup
 where the first mode unfolding of the PARAFAC tensor $[\bm{\mathcal{\hat{\bar{E}}}}]_{(1)} \in \mathbb{C}^{Q_zM_z \times N_y^2N_z^2Q_yM_y}$ shown in \eqref{FORPE_ten} is  given as 
 \begingroup
\setlength{\abovedisplayskip}{0.7ex}
\setlength{\belowdisplayskip}{0.7ex}
\setlength{\abovedisplayshortskip}{0.7ex}
\setlength{\belowdisplayshortskip}{0.7ex}
 \begin{align}
      [\bm{\mathcal{\hat{\bar{E}}}}]_{(1)} = \bm{A}_{\text{z}} \bm{D}_{\bm{\alpha\beta}} \left[\bm{B}_{\text{y}} \diamond  \bm{B}_{\text{z}} \diamond \bm{A}_{\text{y}}\right]^\text{T} .
 \end{align}
 \endgroup
The channel factor $\hat{\bm{A}}_{\text{y}}$ that contains the azimuth angular information of the \ac{BS} and the \ac{UE}, can be therefore estimated
by minimizing the following cost function: 
\begingroup
\setlength{\abovedisplayskip}{0.7ex}
\setlength{\belowdisplayskip}{0.7ex}
\setlength{\abovedisplayshortskip}{0.7ex}
\setlength{\belowdisplayshortskip}{0.7ex}
\begin{align}
      \min_{\boldsymbol{\hat{A}_{\text{y}}}} \left \| [\bm{\mathcal{\hat{\bar{E}}}}]_{(2)} -  \bm{A}_{\text{y}} \bm{D}_{\bm{\alpha\beta}} \left[\bm{B}_{\text{y}} \diamond  \bm{B}_{\text{z}} \diamond \bm{A}_{\text{z}}\right]^\text{T} \right\|  _\text{F}^2, \label{A_y}
\end{align}
\endgroup
where the second mode unfolding of the PARAFAC tensor $ [\bm{\mathcal{\hat{\bar{E}}}}]_{(2)} \in \mathbb{C}^{Q_yM_y \times N_y^2N_z^2Q_zM_z}$ given in \eqref{FORPE_ten} is  given as
\begingroup
\setlength{\abovedisplayskip}{0.7ex}
\setlength{\belowdisplayskip}{0.7ex}
\setlength{\abovedisplayshortskip}{0.7ex}
\setlength{\belowdisplayshortskip}{0.7ex}
 \begin{align}
     [\bm{\mathcal{\hat{\bar{E}}}}]_{(2)} = \bm{A}_{\text{y}} \bm{D}_{\bm{\alpha\beta}} \left[\bm{B}_{\text{y}} \diamond  \bm{B}_{\text{z}} \diamond \bm{A}_{\text{z}}\right]^\text{T} .
 \end{align}
 \endgroup
The channel factor $\hat{\bm{B}}_{\text{z}}$ that contains all the angular information related to the vertical domain of the \ac{BD-RIS} can be estimated by solving the following \ac{LS} cost function:
\begingroup
\setlength{\abovedisplayskip}{0.7ex}
\setlength{\belowdisplayskip}{0.7ex}
\setlength{\abovedisplayshortskip}{0.7ex}
\setlength{\belowdisplayshortskip}{0.7ex}
\begin{align}
      \min_{\boldsymbol{\hat{B}_{\text{z}}}} \left \| [\bm{\mathcal{\hat{\bar{E}}}}]_{(3)} -  \bm{B}_{\text{z}} \bm{D}_{\bm{\alpha\beta}} \left[\bm{B}_{\text{y}} \diamond  \bm{A}_{\text{y}} \diamond \bm{A}_{\text{z}}\right]^\text{T} \right\|  _\text{F}^2, \label{B_z}
\end{align}
\endgroup
 where the third mode unfolding of the PARAFAC tensor $[\bm{\mathcal{\hat{\bar{E}}}}]_{(3)} \in \mathbb{C}^{N_z^2 \times Q_zM_zQ_yM_yN_y^2}$ given in \eqref{FORPE_ten} is  given as
 \begingroup
\setlength{\abovedisplayskip}{0.7ex}
\setlength{\belowdisplayskip}{0.7ex}
\setlength{\abovedisplayshortskip}{0.7ex}
\setlength{\belowdisplayshortskip}{0.7ex}
 \begin{align}
     [\bm{\mathcal{\hat{\bar{E}}}}]_{(3)} = \bm{B}_{\text{z}} \bm{D}_{\bm{\alpha\beta}} \left[\bm{B}_{\text{y}} \diamond  \bm{A}_{\text{y}} \diamond \bm{A}_{\text{z}}\right]^\text{T} .
 \end{align}
 \endgroup
Finally, the channel factor $\hat{\bm{B}}_{\text{y}}$ that contains all the related azimuth angular information of the \ac{BD-RIS} can be estimated by minimizing the following cost function: 
\begingroup
\setlength{\abovedisplayskip}{0.7ex}
\setlength{\belowdisplayskip}{0.7ex}
\setlength{\abovedisplayshortskip}{0.7ex}
\setlength{\belowdisplayshortskip}{0.7ex}
\begin{align}
      \min_{\boldsymbol{\hat{B}_{\text{y}}}} \left \| [\bm{\mathcal{\hat{\bar{E}}}}]_{(4)}  -  \bm{B}_{\text{y}} \bm{D}_{\bm{\alpha\beta}} \left[\bm{B}_{\text{z}} \diamond  \bm{A}_{\text{y}} \diamond \bm{A}_{\text{z}}\right]^\text{T} \right\|  _\text{F}^2, \label{B_y}
\end{align}
\endgroup
 where the fourth mode unfolding of the PARAFAC tensor $[\bm{\mathcal{\hat{\bar{E}}}}]_{(4)} \in \mathbb{C}^{N_y^2 \times Q_zM_zQ_yM_yN_z^2}$ given in \eqref{FORPE_ten} is  given as
 \begingroup
\setlength{\abovedisplayskip}{0.7ex}
\setlength{\belowdisplayskip}{0.7ex}
\setlength{\abovedisplayshortskip}{0.7ex}
\setlength{\belowdisplayshortskip}{0.7ex}
 \begin{align}
     [\bm{\mathcal{\hat{\bar{E}}}}]_{(4)} = \bm{B}_{\text{y}} \bm{D}_{\bm{\alpha\beta}} \left[\bm{B}_{\text{z}} \diamond  \bm{A}_{\text{y}} \diamond \bm{A}_{\text{z}}\right]^\text{T} .
 \end{align}
 \endgroup
Equations \eqref{A_z}, \eqref{A_y}, \eqref{B_z}, \eqref{B_y} will be solved iteratively until they converge by achieving a predefined threshold. 
The steps are shown in Algorithm \ref{algorithm_forpe}.
 \begin{algorithm}[!t]
	\begingroup
	\setlength{\itemsep}{0.2ex}
	\setlength{\parsep}{0pt}
	\setlength{\topsep}{0pt}
	\setlength{\partopsep}{0pt}
	\setlength{\abovedisplayskip}{0.4ex}
	\setlength{\belowdisplayskip}{0.4ex}
	\begin{algorithmic}[1]
		\caption{Fourth-Order PARAFAC Channel Estimation (FORPE)}\label{algorithm_forpe}
		\State \textbf{Inputs}: Estimated composite channel \eqref{E_est}  $\bm{\hat{E}} $, and transform it as tensor $\bm{\mathcal{\hat{\bar{E}}}}$ \eqref{FORPE_ten}, decompose \textit{via} \ac{ALS}:
		\State Set $i=0$. Randomly initialize $\bm{A}_{\text{y}}$, $\bm{B}_{\text{z}}$, $\bm{B}_{\text{y}}$. 
\For{$i = 1:I$}		
		\State Compute an LS estimate of $\bm{A}_{\text{z}}$ as 
  \begin{equation*}  
      \hat{\bm{A}}_{\text{z}} = [\bm{\mathcal{\hat{\bar{E}}}}]_{(1)}\left[\bm{D}_{\bm{\alpha\beta}} \left[\bm{B}_{\text{z}} \diamond  \bm{B}_{\text{y}} \diamond \bm{A}_{\text{y}}\right]^\text{T}\right]^{\dagger}
  \end{equation*}
  	\State Compute an LS estimate of $\bm{A}_{\text{y}}$ as
   \begin{equation*}
      \hat{\bm{A}}_{\text{y}} = [\bm{\mathcal{\hat{\bar{E}}}}]_{(2)}\left[\bm{D}_{\bm{\alpha\beta}} \left[\bm{B}_{\text{z}} \diamond  \bm{B}_{\text{y}} \diamond \bm{A}_{\text{z}}\right]^\text{T}\right]^{\dagger}
  \end{equation*}
   \State Compute an LS estimate of $\bm{B}_{\text{z}}$ as
   \begin{equation*}
      \hat{\bm{B}}_{\text{z}} = [\bm{\mathcal{\hat{\bar{E}}}}]_{(3)}\left[\bm{D}_{\bm{\alpha\beta}} \left[\bm{B}_{\text{y}} \diamond  \bm{A}_{\text{z}} \diamond \bm{A}_{\text{y}}\right]^\text{T}\right]^{\dagger}
  \end{equation*}
   \State Compute an LS estimate of $\bm{B}_{\text{y}}$ as
   \begin{equation*}
      \hat{\bm{B}}_{\text{y}} = [\bm{\mathcal{\hat{\bar{E}}}}]_{(4)}\left[\bm{D}_{\bm{\alpha\beta}} \left[\bm{B}_{\text{z}} \diamond  \bm{A}_{\text{z}} \diamond \bm{A}_{\text{y}}\right]^\text{T}\right]^{\dagger}
  \end{equation*}
   \State Compute an LS estimate of $\hat{\bm{d}}_{\bm{\alpha\beta}}$ as
   \begin{equation*}
     \hat{\bm{d}}_{\bm{\alpha\beta}} = \left[\left( \hat{\bm{B}}_{\text{y}} \diamond \hat{\bm{B}}_{\text{z}}\right) \diamond \left( \hat{\bm{A}}_{\text{y}} \diamond \hat{\bm{A}}_{\text{z}}\right)\right]^\dagger \textrm{vec}\left\{\hat{\bar{\bm{E}}}\right\} 
   \end{equation*}
   \State 
   Compute $\hat{\ma{E}}_{(i)}= \hat{\bm{B}}_{\text{z}} \hat{\bm{D}}_{\bm{\alpha\beta}} \left[ \hat{\bm{B}}_{\text{y}} \diamond  \hat{\bm{A}}_{\text{z}} \diamond \hat{\bm{A}}_{\text{y}} \right]^{\text{T}}$	

   and calculate the error $\epsilon_{(i)} = \big\|[\bm{\mathcal{\hat{\bar{E}}}}]_{(3)}-\hat{\ma{E}}_{(i)} \big\|_{\text{F}}^{2}$,
 
    \State Check convergence and stop if $|\epsilon_{(i)} - \epsilon_{(i-1)}| \leq \eta$.
 \EndFor
		\State Return $\hat{\bm{A}}_{\text{z}} $, $\hat{\bm{A}}_{\text{y}}$, $\hat{\bm{B}}_{\text{z}} $, $\hat{\bm{B}}_{\text{y}}$, and $\hat{\bm{d}}_{\bm{\alpha\beta}}$.
	\end{algorithmic}
	\endgroup
\end{algorithm}
\subsection{Identifiability Conditions}
\textcolor{black}{We now examine identifiability of the \ac{FORTE} and \ac{FORPE} methods. The purpose is to translate the algebraic rank requirements of the \ac{ALS} subproblems into simple dimensional conditions on the numbers of antennas, reflecting elements, and propagation paths. In each \ac{ALS} update, one factor matrix is estimated while the remaining factors and the core are kept fixed. Therefore, the corresponding least-squares solution is unique whenever the regression matrix multiplying the unknown factor has full row rank.}

\textcolor{black}{For the \ac{FORTE} method, the subproblems in \eqref{mode1_forte}, \eqref{mode2_forte}, \eqref{mode3_forte}, and \eqref{mode4_forte} are linear least-squares problems with respect to $\bm{A}_{\text{UE}}$, $\bm{A}_{\text{BS}}$, $\bm{C}_{\text{RIS}}$, and $\bm{B}_{\text{RIS}}$, respectively. To improve readability, let us define the corresponding regression matrices as
\begingroup
\setlength{\abovedisplayskip}{0.5ex}
\setlength{\belowdisplayskip}{0.5ex}
\setlength{\abovedisplayshortskip}{0.5ex}
\setlength{\belowdisplayshortskip}{0.5ex}
\begin{align*}
\bm{Z}_1 &= [\boldsymbol{\mathcal{D}}_{\bm{\alpha\beta}}]_{(1)}
\left[\bm{B}_{\text{RIS}} \otimes \bm{C}_{\text{RIS}} \otimes \bm{A}_{\text{BS}}\right]^\text{T}
\in \mathbb{C}^{L \times MN^2}, \\
\bm{Z}_2 &= [\boldsymbol{\mathcal{D}}_{\bm{\alpha\beta}}]_{(2)}
\left[\bm{B}_{\text{RIS}} \otimes \bm{C}_{\text{RIS}} \otimes \bm{A}_{\text{UE}}\right]^\text{T}
\in \mathbb{C}^{R \times QN^2}, \\
\bm{Z}_3 &= [\boldsymbol{\mathcal{D}}_{\bm{\alpha\beta}}]_{(3)}
\left[\bm{B}_{\text{RIS}} \otimes \bm{A}_{\text{BS}} \otimes \bm{A}_{\text{UE}}\right]^\text{T}
\in \mathbb{C}^{L \times QMN}, \\
\bm{Z}_4 &= [\boldsymbol{\mathcal{D}}_{\bm{\alpha\beta}}]_{(4)}
\left[\bm{C}_{\text{RIS}} \otimes \bm{A}_{\text{BS}} \otimes \bm{A}_{\text{UE}}\right]^\text{T}
\in \mathbb{C}^{R \times QMN}.
\end{align*}
\endgroup
}
\textcolor{black}{Thus, each update is identifiable if the corresponding matrix $\bm{Z}_i$, $i\in\{1,2,3,4\}$, is full row rank, yielding the following necessary dimensional conditions
\begingroup
\setlength{\abovedisplayskip}{0.5ex}
\setlength{\belowdisplayskip}{0.5ex}
\setlength{\abovedisplayshortskip}{0.5ex}
\setlength{\belowdisplayshortskip}{0.5ex}
\begin{align}
     MN^2 \ge L, \,\,  QN^2 \ge R,\nonumber\\
     \,\, QMN \ge L,  \,\, QMN \ge R,
\end{align}
\endgroup
}
\textcolor{black}{Equivalently, the \ac{UE}- and \ac{RIS}-arrival factors require enough observations to resolve the $L$ paths, while the \ac{BS}- and \ac{RIS}-departure factors require enough observations to resolve the $R$ paths. These requirements can be summarized as
\begingroup
\setlength{\abovedisplayskip}{0.8ex}
\setlength{\belowdisplayskip}{0.8ex}
\setlength{\abovedisplayshortskip}{0.8ex}
\setlength{\belowdisplayshortskip}{0.8ex}
\begin{align}
     \max \left( MN^2, QMN\right) \ge L, \,\,    \max \left( QN^2, QMN\right) \ge R.
\end{align}
 \endgroup
}
\textcolor{black}{For the \ac{FORPE} method, identifiability is governed by two complementary requirements. First, since \ac{FORPE} represents the composite channel through a fourth-order \ac{PARAFAC} model of rank $LR$, the decomposition is essentially unique up to the scaling and permutation. A standard sufficient condition is the Kruskal condition, which depends on the $k$-rank $\textrm{Kr}(\cdot)$ of each factor matrix. For our fourth-order tensor channel model, this condition becomes $  \textrm{Kr}(\bm{A}_z) + \textrm{Kr}(\bm{A}_y) + \textrm{Kr}(\bm{B}_z) + \textrm{Kr}(\bm{B}_y) \geq 2LR + 3$}.
\textcolor{black}{This condition is satisfied with high probability when the angular parameters are sufficiently distinct, since the array response matrices tend to have full rank. Second, each update in \ac{FORPE} must be unique. Define the regression matrices in \eqref{A_z}, \eqref{A_y}, \eqref{B_z}, and \eqref{B_y} as
\begingroup
\setlength{\abovedisplayskip}{0.7ex}
\setlength{\belowdisplayskip}{0.7ex}
\setlength{\abovedisplayshortskip}{0.7ex}
\setlength{\belowdisplayshortskip}{0.7ex}
\begin{align*}
\bm{Z}_5 &= \bm{D}_{\bm{\alpha\beta}}
\left[\bm{B}_{\text{z}} \diamond \bm{B}_{\text{y}} \diamond \bm{A}_{\text{y}}\right]^\text{T}
\in \mathbb{C}^{LR \times Q_yM_yN_y^2N_z^2}, \\
\bm{Z}_6 &= \bm{D}_{\bm{\alpha\beta}}
\left[\bm{B}_{\text{z}} \diamond \bm{B}_{\text{y}} \diamond \bm{A}_{\text{z}}\right]^\text{T}
\in \mathbb{C}^{LR \times Q_zM_zN_y^2N_z^2}, \\
\bm{Z}_7 &= \bm{D}_{\bm{\alpha\beta}}
\left[\bm{B}_{\text{y}} \diamond \bm{A}_{\text{z}} \diamond \bm{A}_{\text{y}}\right]^\text{T}
\in \mathbb{C}^{LR \times Q_yM_yQ_zM_zN_y^2}, \\
\bm{Z}_8 &= \bm{D}_{\bm{\alpha\beta}}
\left[\bm{B}_{\text{z}} \diamond \bm{A}_{\text{z}} \diamond \bm{A}_{\text{y}}\right]^\text{T}
\in \mathbb{C}^{LR \times Q_yM_yQ_zM_zN_z^2}.
\end{align*}
\textcolor{black}{Requiring $\bm{Z}_i$, $i\in\{5,6,7,8\}$, to be full row rank yields}
\begin{align}
    Q_yM_yN_y^2N_z^2 \geq LR, \quad Q_zM_zN_y^2N_z^2 \geq LR, \nonumber \\ Q_yM_yQ_zM_zN_y^2 \geq LR, \quad Q_yM_yQ_zM_zN_z^2 \geq LR
\end{align}
\textcolor{black}{which can be compactly written as}
\begin{align}
     \max \left(Q_yM_yN_y^2N_z^2, Q_zM_zN_y^2N_z^2,\right. \nonumber\\
    \left. Q_yM_yQ_zM_zN_y^2,Q_yM_yQ_zM_zN_z^2\right) \ge LR 
\end{align}
\endgroup
}

   \begin{table*}[!t]
\caption{Tensor models adopted under different array geometries at the \ac{BS} and the \ac{UE}.}
\centering
\small
\renewcommand{\arraystretch}{1.15}
\setlength{\tabcolsep}{5pt}
\begin{tabularx}{0.75\textwidth}{@{} p{2.25cm} >{\centering\arraybackslash}p{3.15cm} Y >{\centering\arraybackslash}p{3.25cm} @{}}
\toprule
\textbf{BS/UE} & \textbf{Method} & \textbf{Factor Matrices} & \textbf{Tensor Model} \\
\midrule

\multirow{2}{*}{\ac{UPA}/\ac{UPA}}
& FORTE
& $\left[ \bm{A}_{\mathrm{UE}}, \bm{A}_{\mathrm{BS}}, \bm{C}_{\mathrm{RIS}}, \bm{B}_{\mathrm{RIS}} \right]$
& 4th-order Tucker \\
& FORPE
& $\left[ \bm{A}_{z}, \bm{A}_{y}, \bm{B}_{z}, \bm{B}_{y} \right]$
& 4th-order PARAFAC \\
\midrule

\multirow{2}{*}{\ac{ULA}/\ac{UPA}}
& FORTE
& $\left[ \bm{A}_{\mathrm{UE}}, \bm{A}_{\mathrm{BS}}, \bm{C}_{\mathrm{RIS}}, \bm{B}_{\mathrm{RIS}} \right]$
& 4th-order Tucker \\
& FORPE
& $\left[ \bm{A}_{z}, \bm{A}^{y}_{\mathrm{UE}}\bm{\Psi}, \bm{B}_{z}, \bm{B}_{y} \right]$
& Constrained PARAFAC \\
\midrule

\multirow{2}{*}{\ac{UPA}/\ac{ULA}}
& FORTE
& $\left[ \bm{A}_{\mathrm{UE}}, \bm{A}_{\mathrm{BS}}, \bm{C}_{\mathrm{RIS}}, \bm{B}_{\mathrm{RIS}} \right]$
& 4th-order Tucker \\
& FORPE
& $\left[ \bm{A}_{z}, \bm{A}^{y}_{\mathrm{BS}}\bm{\Phi}, \bm{B}_{z}, \bm{B}_{y} \right]$
& Constrained PARAFAC \\
\midrule

\multirow{2}{*}{\ac{ULA}/\ac{ULA}}
& FORTE
& $\left[ \bm{A}_{\mathrm{UE}}, \bm{A}_{\mathrm{BS}}, \bm{C}_{\mathrm{RIS}}, \bm{B}_{\mathrm{RIS}} \right]$
& 4th-order Tucker \\
& FORPE
& $\left[ \bm{A}_{z}, \bm{B}_{z}, \bm{B}_{y} \right]$
& 3rd-order PARAFAC \\
\bottomrule
\end{tabularx}
\label{tb:array}
\end{table*}

\subsection{Computational Complexity}
\textcolor{black}{We evaluate the computational complexity by counting the dominant operations required by each estimator. 
The initial unstructured \ac{LS} estimate of the composite channel follows from \eqref{filtered_signal} and \eqref{E_est}; its dominant cost is written as $\mathcal{O}(Q^3T^2K^2MN^2)$ since the inversion and multiplication involve the pilot, combiner, and scattering dimensions jointly.}

\textcolor{black}{For \ac{FORTE}, Algorithm~\ref{algorithm_forte} updates four factor matrices per iteration. The four associated regression matrices have dimensions $L\times MN^2$, $R\times QN^2$, $L\times QMN$, and $R\times QMN$, respectively. 
This gives the per-iteration costs $\mathcal{O}(L^2MN^2)$, $\mathcal{O}(R^2QN^2)$, $\mathcal{O}(L^2QMN)$, and $\mathcal{O}(R^2QMN)$. Hence, after $J$ \ac{ALS} iterations, the total dominant complexity of \ac{FORTE} is $\mathcal{O} \left(J \left(  L^2MN^2+R^2QN^2+L^2QMN+R^2QMN\right)\right)$}.
\textcolor{black}{For \ac{FORPE}, the same reasoning is applied to the four \ac{PARAFAC} updates in Algorithm~\ref{algorithm_forpe}. Since the rank of the fully structured model is $LR$, 
we have}
$\mathcal{O}\big(
P\big(
L^2R^2Q_yM_yN_y^2N_z^2
\allowbreak + L^2R^2Q_zM_zN_y^2N_z^2
\allowbreak + L^2R^2QM N_y^2
\allowbreak + L^2R^2QM N_z^2
\big)
\big)$, \textcolor{black}{where $P$ is the number of \ac{ALS} iterations required by \ac{FORPE}. Finally, for a fully connected architecture, the competing \ac{BTKF} estimator in \cite{Andre_Almeida_2025} first obtains an unstructured \ac{LS} estimate and then applies a rank-one Kronecker-factorization step to recover the channel factors. The former contributes $\mathcal{O}(Q^3T^2K^2MN^2)$, while the latter is dominated by the rank-one approximation of the rearranged composite channel matrix and contributes $\mathcal{O}(QMN^2)$. Thus, the overall dominant complexity is $\mathcal{O}(Q^3T^2K^2MN^2 + QMN^2)$.}

\begin{table*}[!t]
\caption{\textcolor{black}{Summary of identifiability, complexity, and modeling tradeoffs of the proposed tensor estimators.}}
\label{tb:ident_complexity_tradeoff}
\centering
\scriptsize
\begingroup\color{black}
\renewcommand{\arraystretch}{1.25}
\setlength{\tabcolsep}{2.5pt}
\begin{tabularx}{\textwidth}{@{}p{1.35cm} p{3.0cm} p{4.4cm} p{4.1cm} X@{}}
\toprule
\textbf{Method} & \textbf{Tensor model} & \textbf{Identifiability requirement} & \textbf{Dominant complexity} & \textbf{Main tradeoff} \\
\midrule
\ac{FORTE} &
Fourth-order Tucker with factors $\bm{A}_{\mathrm{UE}}$, $\bm{A}_{\mathrm{BS}}$, $\bm{C}_{\mathrm{RIS}}$, and $\bm{B}_{\mathrm{RIS}}$. &
\(\begin{gathered}
\max(MN^2,QMN)\ge L,\\
\max(QN^2,QMN)\ge R.
\end{gathered}\) &
\(\begin{gathered}
\mathcal{O}\!\left(J( L^2MN^2+R^2QN^2\right.\\
\left.+L^2QMN+R^2QMN)\right)
\end{gathered}\) &
More compact and robust; provides lower representation burden; does not guarantee direct unique recovery of all directional factors but identifies factor subspaces. \\
\midrule
\ac{FORPE} &
Fourth-order \ac{PARAFAC} with horizontal and vertical factors $\bm{A}_z$, $\bm{A}_y$, $\bm{B}_z$, and $\bm{B}_y$. &
\(\begin{gathered}
\max\big(Q_yM_yN_y^2N_z^2,\\
Q_zM_zN_y^2N_z^2,\\
Q_yM_yQ_zM_zN_y^2,\\
Q_yM_yQ_zM_zN_z^2\big)\ge LR.
\end{gathered}\) &
\(\begin{gathered}
\mathcal{O}\!\left(P L^2R^2( Q_yM_yN_y^2N_z^2\right.\\
+Q_zM_zN_y^2N_z^2\\
\left.+QMN_y^2+QMN_z^2)\right)
\end{gathered}\) &
Finer deconstruction with stronger essential uniqueness and direct directional-factor recovery; higher representation and arithmetic complexity, and more sensitivity to noise. \\
\bottomrule
\end{tabularx}
\endgroup
\end{table*}

\textcolor{black}{Table~\ref{tb:ident_complexity_tradeoff} summarizes identifiability, complexity, and tradeofs. While \ac{FORTE} favors compactness and numerical robustness, \ac{FORPE} favors a finer and unique directional deconstruction of the composite channel at the cost of a larger parameter space and higher per-iteration complexity.}

\subsection{Representation complexity of \ac{FORTE} and \ac{FORPE} methods}
\textcolor{black}{We now compare the representation complexity of the proposed tensor models, i.e., the number of scalar coefficients required to parameterize the composite channel representation before the iterative estimation is carried out. 
Lower representation complexity generally means fewer parameters must be estimated from the same training data, improving robustness in noisy and finite-sample regimes.}

\textcolor{black}{For \ac{FORTE}, the fourth-order Tucker representation in \eqref{Tuker_4} contains four factor matrices and one core tensor. The factor matrices $\bm{A}_{\text{UE}}\in\mathbb{C}^{Q\times L}$, $\bm{A}_{\text{BS}}\in\mathbb{C}^{M\times R}$, $\bm{C}_{\text{RIS}}\in\mathbb{C}^{N\times L}$, and $\bm{B}_{\text{RIS}}\in\mathbb{C}^{N\times R}$ contribute $QL$, $MR$, $NL$, and $NR$ coefficients, respectively. The Tucker core $\boldsymbol{\mathcal{D}}_{\bm{\alpha\beta}}$ contributes $L^2R^2$ coefficients in the general representation. Hence, the total representation complexity of \ac{FORTE} is
\begin{align}
    \textrm{C}_{\textrm{FORTE}}=\left( QL + MR +NL+ NR\right) + L^2R^2.
\end{align}
}
\textcolor{black}{This count is a conservative upper bound for the Tucker representation. If the Kronecker-diagonal structure $\bm{D}_{\bm{\alpha\beta}}=\bm{D}(\bm{\alpha})\otimes\bm{D}(\bm{\beta})$ is explicitly enforced, the number of independent gain coefficients is reduced from $L^2R^2$ to $LR$, corroborating the compactness of \ac{FORTE}. We keep the general $L^2R^2$ count to remain consistent with the unconstrained Tucker-core update used in the \ac{ALS} implementation.}

\textcolor{black}{For \ac{FORPE}, the fourth-order \ac{PARAFAC} model explicitly separates the horizontal and vertical spatial responses. The factor matrices have $LR$ columns, corresponding to all combinations of the $L$ \ac{BD-RIS}--\ac{UE} paths and the $R$ \ac{BS}--\ac{BD-RIS} paths. The factors associated with the $z$- and $y$-dimensions of the terminal arrays contribute $Q_zM_zLR$ and $Q_yM_yLR$ coefficients, while the two \ac{BD-RIS} spatial dimensions contribute $N_z^2LR$ and $N_y^2LR$ coefficients. The path-gain vector adds $LR$ coefficients. Therefore, the total representation complexity of \ac{FORPE} is
\begin{align}
   \textrm{C}_{\textrm{FORPE}}=\left( Q_zM_z + Q_yM_y+N_z^2+N_y^2\right)LR + LR.
\end{align}
}
\textcolor{black}{As an example, consider the setting in Table~\ref{tb:simul}, with $M=16$, $Q=4$, $N=64$, and $L=R=2$. For planar arrays, we use $M_zM_y=16$, $Q_zQ_y=4$, and $N_zN_y=64$, yielding $M_z=M_y=4$, $Q_z=Q_y=2$, and $N_z=N_y=8$. Under these values, \ac{FORTE} requires $QL+MR+NL+NR+L^2R^2=312$ coefficients, whereas \ac{FORPE} requires $\left(Q_zM_z+Q_yM_y+N_z^2+N_y^2\right)LR+LR=580$ coefficients. Thus, \ac{FORPE} uses approximately $1.86$ times more parameters than \ac{FORTE} in this setting.}
\textcolor{black}{Generally, the relative representation burden can be expressed as $\eta_{\text{rep}}=\textrm{C}_{\textrm{FORPE}}/\textrm{C}_{\textrm{FORTE}}$.}
\textcolor{black}{Note that values $\eta_{\text{rep}}>1$ indicate that \ac{FORPE} uses a larger number of model coefficients than \ac{FORTE}. The ratio increases with the explicit two-dimensional factorization of the array responses, especially through the $N_z^2LR$ and $N_y^2LR$ terms associated with the two \ac{BD-RIS} dimensions.}

\textcolor{black}{\textit{Remark 1}: This difference explains part of the behavior observed in the numerical results. The \ac{FORTE} model is more compact because it keeps the main channel factors grouped by physical links, namely the UE responses and the two \ac{BD-RIS} responses. This grouping reduces the number of free factors and makes the corresponding \ac{ALS} updates less sensitive to noise. By contrast, \ac{FORPE} imposes a more detailed parametric structure by decomposing the array responses into horizontal and vertical components, thereby increasing the number of coefficients and potentially making parameter estimation more sensitive to errors, especially at low \ac{SNR} or with limited training.}
\textcolor{black}{The two models therefore involve a tradeoff. \ac{FORTE} provides a lower-dimensional and numerically robust representation, which is beneficial for reconstructing the composite channel. However, its Tucker structure identifies the factor subspaces rather than all individual physical factors directly. \ac{FORPE}, on the other hand, has a larger representation complexity, but its \ac{PARAFAC} structure can provide essentially unique factor estimates under the identifiability conditions discussed above. This uniqueness is useful when the goal is not only to reconstruct the composite channel but also to extract the underlying angular parameters.}

\subsection{Special Cases for Different Array Structures}
Table \ref{tb:array} summarizes how the proposed tensor models specialize when the \ac{BS} and the \ac{UE} employ different array geometries. The purpose of this table is to clarify how the factor matrices and the associated tensor model change when moving from the general \ac{UPA}/\ac{UPA} setting to mixed \ac{ULA}/\ac{UPA}, \ac{UPA}/\ac{ULA}, and \ac{ULA}/\ac{ULA} configurations. In all cases, the \ac{FORTE} model preserves the same fourth-order Tucker structure, since it operates at the level of the main channel factors $\bm{A}_{\text{UE}}$, $\bm{A}_{\text{BS}}$, $\bm{C}_{\text{RIS}}$, and $\bm{B}_{\text{RIS}}$, independently of whether the terminal arrays are one-dimensional or two-dimensional. By contrast, the \ac{FORPE} model is affected by the array geometry because it explicitly decomposes the steering matrices into horizontal and vertical components. 

Hence, for the \ac{UPA}/\ac{UPA} case, \ac{FORPE} retains its full fourth-order \ac{PARAFAC} form with factor matrices $\left[\bm{A}_{\text{z}},\bm{A}_{\text{y}},\bm{B}_{\text{z}},\bm{B}_{\text{y}}\right]$. When one side uses a \ac{ULA}, the missing spatial dimension is absorbed through the constraint matrices $\bm{\Psi} = \bm{1}_R^\text{T} \otimes \bm{I}_L$ and $\bm{\Phi} = \bm{I}_R \otimes \bm{1}_L^\text{T}$, yielding constrained \ac{PARAFAC} models \cite{favier2014overview} for the \ac{ULA}/\ac{UPA} and \ac{UPA}/\ac{ULA} cases. Finally, when both the \ac{BS} and the \ac{UE} are equipped with \ac{ULA}s, and assuming that each \ac{ULA} is oriented along the $z$-dimension, the \ac{FORPE} representation further reduces to a third-order \ac{PARAFAC} model since one spatial dimension is absent at both ends. Table \ref{tb:array} makes explicit that \ac{FORTE} is structurally robust across array configurations, whereas \ac{FORPE} adapts its factorization according to the available spatial dimensions and becomes constrained or reduced whenever one or both terminal arrays are one-dimensional.

\section{\ac{TenFormer} based Beamforming Design}
In this section, we show how the estimated composite channel can be exploited to design joint beamforming in \ac{BD-RIS}-assisted \ac{MIMO} systems. We formulate a tractable design framework for the precoder, combiner, and \ac{BD-RIS} scattering matrix consistent with the proposed tensor-based channel models for spectral efficiency maximization.

Consider a \ac{MIMO} communication system, where the \ac{BS} is communicating with a single \ac{UE} \textit{via} \ac{BD-RIS} as shown in Figure \ref{fig:system_model}. Assuming that the \ac{BS} transmits a symbol vector $\bm{s}\in \bb{C}^{N_s \times 1}$, where $N_s$ is the data streams given that $N_s \leq \min \left\{ R,L \right\}$ and $\bb{E}\left\{\bm{s}\bm{s}^{\text{T}}\right\} = \bm{I}_{N_s}$. The estimated data symbol vector at the \ac{UE} is given as:
\begingroup
\setlength{\abovedisplayskip}{0.5ex}
\setlength{\belowdisplayskip}{0.5ex}
\setlength{\abovedisplayshortskip}{0.5ex}
\setlength{\belowdisplayshortskip}{0.5ex}
\begin{align}
    \hat{\bm{s}} =  \bm{W}^\text{H} \bm{P} \bm{F} \bm{s} + \bm{W}^\text{H}\bm{v}^\prime \in \mathbb{C}^{N_s \times 1}, \label{beamforming}
\end{align} 
\endgroup
where $\bm{P} = \bm{G} \bm{\Theta} \bm{H} \in \mathbb{C}^{Q \times M}$ is the so called cascaded channel.  $\bm{W}\in \mathbb{C}^{Q \times N_s}$ is the combiner, and $\bm{F}\in \mathbb{C}^{M \times N_s}$ is the transmit precoder. $\bm{v}=\bm{W}^\text{H}\bm{v}^\prime  \sim \mathcal{CN}\left(\bm{0},\sigma_n^2\bm{I}_{N_s \times 1}\right)$ denotes the filtered noise. The joint transceiver and BD-RIS design problem to maximize the spectral efficiency is given as
\begin{align}
    \max_{\bm{W,F,\Theta}} \log_2 \left( \left| \bm{I}_{N_s} + \frac{\bm{W}^\text{H}\bm{PF}\bm{F}^\text{H}\bm{P}^\text{H}\bm{W}}{\sigma_n^2\bm{W}^\text{H}\bm{W}}\right|\right) \nonumber \\
    \text{subject to} \;\; \bm{\Theta}^\text{H}\bm{\Theta} = \bm{I}_N, \text{and} \;\; \text{tr}\left\{ \bm{F}\bm{F}^\text{H}\right\}\leq P_T \label{cost_fuc_spec_eff}
\end{align}
where $P_T$ is the total transmit power. The non-convex \ac{BD}-\ac{RIS} constraints render the above optimization problem difficult to solve. To render the problem in \eqref{cost_fuc_spec_eff} amenable to tensor-based processing, we adopt a surrogate formulation in which the achievable-rate objective is approximated by the Frobenius norm of the effective end-to-end channel. This yields the following simplified optimization problem:
\begin{align}
    &\max_{\bm{W,F,\Theta}}  \left\| \bm{W}^\text{H}\bm{G}\bm{\Theta}\bm{H}\bm{F}\right\|_\text{F}^2 \nonumber \\
    \text{subject to} \;\; &\bm{\Theta}^\text{H}\bm{\Theta} = \bm{I}_N, \text{and} \;\; \text{tr}\left\{ \bm{F}\bm{F}^\text{H}\right\}\leq P_T .
\end{align}
Compared with the original rate-maximization problem, this surrogate formulation is more tractable and exhibits a multilinear structure that can be efficiently exploited by tensor optimization methods. Consequently, the precoder, combiner, and BD-RIS can be optimized in parallel within a unified tensor framework.
Applying $\textrm{vec}\{.\}$ operator to the above equation leads to the following optimization problem:
\begin{align}
    \max_{\bm{W,F,\Theta}}  \left\| \left( \bm{F}^\text{T} \otimes \bm{W}^\text{H}\right) \textrm{vec}\left( \bm{G}\bm{\Theta} \bm{H}\right) \right\|_2^2 \nonumber \\
    \text{subject to} \;\; \bm{\Theta}^\text{H}\bm{\Theta} = \bm{I}_N, \text{and} \;\; \text{tr}\left\{ \bm{F}\bm{F}^\text{H}\right\}\leq P_T 
\end{align}
Assuming that $\bm{\Theta} = \bm{\Theta}_G \bm{\Theta}_H^\text{T}$, the problem can be recast as
\begin{align}
   \notag& \max_{\bm{W,F,\bm{\Theta}_G,\bm{\Theta}_H}} \hspace{-0.1cm} \left\| \left( \bm{F}^\text{T} \otimes \bm{W}^\text{H}\right)\hspace{-0.1cm}\left(\ma{H}^{\text{T}} \otimes \ma{G}\right)\hspace{-0.1cm}\left(\bm{\Theta}_H\otimes \bm{\Theta}_G\right) \textrm{vec}\left( \ma{I}_N\right) \right\|_2^2 \nonumber \\
   \label{eq:problem_beam_sep_RIS}& \text{subject to} \;\; \bm{\Theta}^\text{H}\bm{\Theta} = \bm{I}_N, \text{and} \;\; \text{tr}\left\{ \bm{F}\bm{F}^\text{H}\right\}\leq P_T. 
\end{align}
Defining the equivalent channel as $\ma{T}  = ( \bm{F}^\text{T} \otimes \bm{W}^\text{H})(\ma{H}^{\text{T}} \otimes \ma{G}) (\bm{\Theta}_H\otimes \bm{\Theta}_G)$, and noting that $\textrm{vec}\left( \ma{I}_N\right)$ is fixed, the optimization can be focused on $\ma{T}$. Assuming the singular value decompositions (SVDs) $\ma{H}=\ma{U}_{H}\ma{\Sigma}_H\ma{V}_{H}^{\text{H}} \in \bb{C}^{N \times M}$ and $\ma{G} = \ma{U}_{G}\ma{\Sigma}_{G}\ma{V}_{G}^{\text{H}} \in \bb{C}^{Q \times N}$, we can write $\ma{H}_e$ as
\begin{align*}
    \ma{T} &= ( \bm{F}^\text{T} \otimes \bm{W}^\text{H})(\ma{V}_{H}^{*}\ma{\Sigma}_H\ma{U}_H^{\text{T}} \otimes \ma{U}_{G}\ma{\Sigma}_{G}\ma{V}_{G}^{\text{H}}  ) (\bm{\Theta}_H\otimes \bm{\Theta}_G) \\
    &=( \bm{F}^\text{T}\ma{V}_{H}^{*} \otimes \ma{W}^{\text{H}}\ma{U}_G) (\ma{\Sigma}_H \otimes \ma{\Sigma}_H)(\ma{U}_{H}^{\text{T}}\ma{\Theta}_H \otimes \ma{V}_{G}^{\text{H}}\ma{\Theta}_G ),
\end{align*}
from which a solution to \eqref{eq:problem_beam_sep_RIS} is obtained by setting $\ma{F}^{\text{T}} = \ma{V}^{\text{T}}_H$, $\ma{W}^{\text{H}} = \ma{U}^{\text{H}}_G$, $\ma{\Theta}_H = \ma{U}^*_H$, and $\ma{\Theta}_G = \ma{V}_{G}$. In practice, however, this requires separate estimates of $\ma{G}$ and $\ma{H}$. 


 Consider the composite channel $\ma{E}=\ma{H}^{\text{T}} \otimes \ma{G} \in \bb{C}^{QM \times N^2}$, reshaped into a tensor $\ten{E} \in \bb{C}^{Q \times M \times N \times N}$. To illustrate this operation, each  slice of $\ten{E}$, i.e., $\ten{E}_{.,.,n_1,n_2}$, is given by
\begin{align} \label{tensor_channel}
    \ten{E}_{.,.,n_1,n_2} = \ma{G}_{.,n_1}\ma{H}_{n_2,.}\in \bb{C}^{Q \times M},
\end{align}
for $\{n_1,n_2\} = \{1,\ldots,N\}$. Using this tensor representation, we can recast the problem in \eqref{eq:problem_beam_sep_RIS} as 
\begingroup
\setlength{\abovedisplayskip}{0.5ex}
\setlength{\belowdisplayskip}{0.5ex}
\setlength{\abovedisplayshortskip}{0.5ex}
\setlength{\belowdisplayshortskip}{0.5ex}
\begin{align}
   \notag& \max_{\bm{W,F,\bm{\Theta}_G,\bm{\Theta}_H}} \hspace{-0.1cm} \left\|  \ten{E} \times_1 \ma{W}^{\text{H}} \times_2 \ma{F}^{\text{T}} \times_3 \ma{\Theta}_G \times_4 \ma{\Theta}_H \right\|^2_{\text{F}} \nonumber \\
   \label{eq:problem_tensor}& \text{subject to} \;\; \bm{\Theta}^\text{H}\bm{\Theta} = \bm{I}_N, \text{and} \;\; \text{tr}\left\{ \bm{F}\bm{F}^\text{H}\right\}\leq P_T. 
\end{align}
\endgroup
where $\textrm{vec}\left( \ma{I}_N\right)$ is omitted because it is constant, and the objective is now to maximize the Frobenius norm of the equivalent channel tensor. To solve \eqref{eq:problem_tensor}, we employ the HOSVD, which exploits each unfolding of $\ten{E}$ independently. Assuming that the SVD of $\nmode{\ten{E}}{n}$ is given by $\ma{U}^{(n)}\ma{\Sigma}^{(n)}\ma{V}^{(n)\text{H}}$ for $n= \{1,2,3,4\}$, we design the beamforming matrices as
\begin{align}
    \ma{W} &= \ma{U}^{(1)}_{;,1:R_s} \in \bb{C}^{Q \times R_s}\, \quad \ma{F} = \ma{U}^{(2)}_{:,1:R_s} \in \bb{C}^{M \times R_s}\\
    \ma{\Theta}_G &=\ma{U}^{(3)} \in \bb{C}^{N \times N}\,\quad \ma{\Theta}_H =\ma{U}^{(4)} \in \bb{C}^{N \times N},
\end{align}
while the optimum scattering matrix is given by $\ma{\Theta} = \ma{\Theta}_G \ma{\Theta}_H^{\text{T}}$.

Although this procedure yields the same performance as the matrix-based solution of \eqref{eq:problem_beam_sep_RIS}, the proposed tensor-based approach offers two main advantages. First, it does not require separate channel estimation, since the beamformers can be optimized directly from the estimated composite channel. Second, it relies on four independent SVDs, which can be computed in parallel. The proposed \ac{TenFormer} beamforming design is summarized in Algorithm \ref{algorithm_tenformer}.

An important advantage of the proposed \ac{TenFormer} design is that it operates directly on the tensorized composite channel, thereby avoiding the additional channel-separation stage. Using the standard \ac{SVD} complexity result $\mathcal{O}(\max\{m,n\}\min\{m,n\}^{2})$ for an $m \times n$ matrix \cite{GolubVanLoan2013}, and noting that this reduces to $\mathcal{O}(m^{2}n)$ when $m<n$, TenFormer requires four SVDs, but these decompositions are independent and can be computed in parallel. Hence, the processing delay is determined by the most expensive unfoldings, which are $\nmode{\ten{E}}{3}$ and $\nmode{\ten{E}}{4}$, namely the third and fourth mode unfoldings, resulting in a dominant complexity of order \( \mathcal{O}(M^{2}N^{3}) \) when \( M=Q \). 

By contrast, a beamforming strategy based on separate channel knowledge, such as in \cite{Emil_ozlem_2025}, must first recover these channels from the estimated composite channel, i.e., decoupling $\ma{G}$ and $\ma{H}$ from $\ma{H}^{\text{T}} \otimes \ma{G}$. This additional step can be implemented, for instance, via a classical least-squares Kronecker factorization \cite{VanLoan}. Under the assumption \( M=Q \), this separation stage requires a rank-one SVD of a rearranged \( MN \times MN \) matrix, which incurs a complexity of order \( \mathcal{O}(M^{3}N^{3}) \), dominating the subsequent beamforming stage. Therefore, the proposed TenFormer remains computationally more attractive in practice, since it bypasses the cost of the channel-separation block and directly exploits the multidimensional structure of the combined.

\begin{algorithm}[!t]
\begingroup
\setlength{\itemsep}{0.2ex}
\setlength{\parsep}{0pt}
\setlength{\topsep}{0pt}
\setlength{\partopsep}{0pt}
\setlength{\abovedisplayskip}{0.4ex}
\setlength{\belowdisplayskip}{0.4ex}
\begin{algorithmic}[1]
\caption{\ac{TenFormer} Beamforming Design}\label{algorithm_tenformer}
\State \textbf{Input}: Tensorized composite channel $\ten{E} \in \bb{C}^{Q \times M \times N \times N}$ in \eqref{tensor_channel}, number of streams $R_s$.
\State Form the mode-$n$ unfoldings $\nmode{\ten{E}}{n}$ for $n \in \{1,2,3,4\}$.
\For{$n = 1:4$}
    \State Compute the SVD of the $n$th unfolding:
    \begin{equation*}
        \nmode{\ten{E}}{n} = \ma{U}^{(n)}\ma{\Sigma}^{(n)}\ma{V}^{(n)\text{H}}.
    \end{equation*}
\EndFor
\State Define the receive combiner from the dominant left singular subspace of $\nmode{\ten{E}}{1}$:
\begin{equation*}
    \ma{W} = \ma{U}^{(1)}_{:,1:R_s} \in \mathbb{C}^{Q\times R_s}.
\end{equation*}
\State Define the transmit precoder from the dominant left singular subspace of $\nmode{\ten{E}}{2}$:
\begin{equation*}
    \ma{F} = \ma{U}^{(2)}_{:,1:R_s} \in \mathbb{C}^{M \times R_s}.
\end{equation*}
\State Set the two \ac{BD-RIS} factors from the dominant eigenmodes from $\nmode{\ten{E}}{3}$ and $\nmode{\ten{E}}{4}$, repectively:
\begin{equation*}
    \ma{\Theta}_G = \ma{U}^{(3)} \in \mathbb{C}^{N \times N}, \qquad \ma{\Theta}_H = \ma{U}^{(4)} \in \mathbb{C}^{N \times N}.
\end{equation*}
\State Construct the \ac{BD-RIS} scattering matrix:
\begin{equation*}
    \ma{\Theta} = \ma{\Theta}_G\ma{\Theta}_H^{\text{T}} \in \mathbb{C}^{N \times N}.
\end{equation*}
\State Form the effective beamformed tensor/channel:
\State \textbf{Output}:  $\ma{F}$,   $\ma{W}$,  $\ma{\Theta}$ 
\end{algorithmic}
\endgroup
\end{algorithm}

\vspace{-2ex}
\section{ Cramér--Rao Lower Bound (CRLB)}
\textcolor{black}{In this section, we derive the \ac{CRLB} for the unstructured composite channel in \ac{BD-RIS}-assisted communication systems. Starting from \eqref{main_eq_k}, vectorization gives $\bm{y}_k=(\bm{X}^\text{T}\otimes\bm{I}_Q)\textrm{vec}(\bm{G}\bm{\Omega}_k\bm{H})+\bm{n}_k$, where $\bm{n}_k\sim\mathcal{CN}(\bm{0}_{QT\times 1},\sigma_n^2\bm{I}_{QT})$. Using $\textrm{vec}(\bm{A}\bm{B}\bm{C})=(\bm{C}^\text{T}\otimes\bm{A})\textrm{vec}(\bm{B})$, this becomes $\bm{y}_k=(\bm{X}^\text{T}\otimes\bm{I}_Q)(\bm{H}^\text{T}\otimes\bm{G})\textrm{vec}(\bm{\Omega}_k)+\bm{n}_k$, or equivalently $\bm{y}_k=(\textrm{vec}(\bm{\Omega}_k)^\text{T}\otimes\bm{X}^\text{T}\otimes\bm{I}_Q)\textrm{vec}(\bm{H}^\text{T}\otimes\bm{G})+\bm{n}_k$. Collecting all $K$ received vectors in $\bm{y}=[\bm{y}_1^\text{T},\dots,\bm{y}_K^\text{T}]^\text{T}\in\mathbb{C}^{QTK\times 1}$, the observation model can be written compactly as}
\begingroup
\setlength{\abovedisplayskip}{0.7ex}
\setlength{\belowdisplayskip}{0.7ex}
\setlength{\abovedisplayshortskip}{0.7ex}
\setlength{\belowdisplayshortskip}{0.7ex}
\begin{align}\label{main_eq_crlb}
    \bm{y} = \left( \bm{\Omega}^\text{T} \otimes \bm{X}^\text{T} \otimes \bm{I}_Q\right) \textrm{vec} \left( \bm{H}^\text{T} \otimes \bm{G}\right) + \bm{n},
\end{align}
\endgroup
\textcolor{black}{where $\bm{\Omega}=[\textrm{vec}(\bm{\Omega}_1),\dots,\textrm{vec}(\bm{\Omega}_K)]\in\mathbb{C}^{N^2\times K}$ and $\bm{n}=[\bm{n}_1^\text{T},\dots,\bm{n}_K^\text{T}]^\text{T}\in\mathbb{C}^{QTK\times 1}$ with $\bm{n}\sim\mathcal{CN}(\bm{0},\sigma_n^2\bm{I}_{QTK})$. Let $\bm{\eta}=\textrm{vec}(\bm{H}^\text{T}\otimes\bm{G})\in\mathbb{C}^{QMN^2}$ be the unknown parameter vector and let $\hat{\bm{\eta}}$ be an unbiased estimator. Then, $\mathbb{E}[\lVert\bm{\eta}-\hat{\bm{\eta}}\rVert_2^2]\geq\mathrm{tr}\{\mathrm{CRLB}(\bm{\eta})\}$, where the \ac{CRLB} matrix is the inverse of the \ac{FIM}. Defining $\bm{V}=\bm{\Omega}^\text{T}\otimes\bm{X}^\text{T}\otimes\bm{I}_Q$, \eqref{main_eq_crlb} becomes $\bm{y}=\bm{V}\bm{\eta}+\bm{n}$. Hence, $\bm{y}\sim\mathcal{CN}(\bm{\mu},\bm{R})$,}
where $\bm{\mu} = \bm{V}\bm{\eta}$ and $\bm{R} = \sigma_n^2\bm{I}_{QTK}$. We introduce the augmented real parameter vector $    \bm{\eta}_c = \left[\bar{\bm{\eta}}^\text{T} \,\, \tilde{\bm{\eta}}^\text{T}\right]^\text{T}$,
where $\bar{\bm{\eta}} = \mathrm{Re}\left\{ \bm{\eta}\right\}$ and $\tilde{\bm{\eta}} = \mathrm{Im}\left\{ \bm{\eta}\right\}$. Since the covariance is independent of the unknown parameters, the second term of the Slepian--Bangs formula vanishes, and the \ac{FIM} of $\bm{\eta}_c$ becomes
\begin{equation}
\bm{F}\left(\bm{\eta}_c\right)
=\dfrac{2}{\sigma_n^2}
\begin{bmatrix}
\mathrm{Re}\!\left\{\bm{V}^{\mathrm{H}}\bm{V}\right\}
&
-\mathrm{Im}\!\left\{\bm{V}^{\mathrm{H}}\bm{V}\right\}
\\[6pt]
\mathrm{Im}\!\left\{\bm{V}^{\mathrm{H}}\bm{V}\right\}
&
\mathrm{Re}\!\left\{\bm{V}^{\mathrm{H}}\bm{V}\right\}
\end{bmatrix}.
\end{equation}
\textcolor{black}{Now define $\bar{\bm{Z}}=\mathrm{Re}\{\bm{V}^\text{H}\bm{V}\}$ and $\tilde{\bm{Z}}=\mathrm{Im}\{\bm{V}^\text{H}\bm{V}\}$. Using standard block-matrix inversion formulas, we obtain}
\begin{align}
    \mathrm{tr}\left( \mathrm{CRLB}\left( \bar{\bm{\eta}}\right)\right)
    = \frac{\sigma_n^2}{2} \, \mathrm{tr} \left\{ \left( \bar{\bm{Z}} + \tilde{\bm{Z}} \, \bar{\bm{Z}}^{-1} \, \tilde{\bm{Z}} \right)^{-1} \right\}.
\end{align}
Next, using the orthogonality of the normalized pilot and scattering matrices, $ \bm{X}^\ast\bm{X}^\text{T} = \bm{I}_M, \,\,\,\, \bm{\Omega}^\ast\bm{\Omega}^\text{T} = \bm{I}_{N^2}$,
\textcolor{black}{we obtain $\bm{V}^\text{H}\bm{V}=(\bm{\Omega}^\ast\bm{\Omega}^\text{T})\otimes(\bm{X}^\ast\bm{X}^\text{T})\otimes\bm{I}_Q=\bm{I}_{QMN^2}$. Therefore, $\bar{\bm{Z}}=\bm{I}_{QMN^2}$ and $\tilde{\bm{Z}}=\bm{0}$,}
and the \ac{CRLB} matrices for the real and imaginary parts reduce to $    \mathrm{CRLB}\left( \bar{\bm{\eta}}\right) = \frac{\sigma_n^2}{2}\bm{I}_{QMN^2}$ and $    \mathrm{CRLB}\left( \tilde{\bm{\eta}}\right) = \frac{\sigma_n^2}{2}\bm{I}_{QMN^2}$.
\textcolor{black}{Hence, the augmented real parameter vector satisfies $\mathrm{CRLB}(\bm{\eta}_c)=\frac{\sigma_n^2}{2}\bm{I}_{2QMN^2}$. Thus, the total mean-square error of any unbiased estimator obeys $\mathbb{E}[\|\bm{\eta}_c-\hat{\bm{\eta}}_c\|_2^2]\geq\mathrm{tr}\{\mathrm{CRLB}(\bm{\eta}_c)\}=\sigma_n^2 QMN^2$.}

\renewcommand\baselinestretch{.8}

{\color{black}
\section{Simulation Results}
\begin{table}[t]
	\caption{Simulation parameters}
	\centering
	\footnotesize
	\begin{tabular}{|l|c|}
		\hline
		UPA size (BS) & $M_y M_z = 4 \times 4 = 16$ \\
		\hline
		UPA size (UE) & $Q_y Q_z = 2 \times 2 = 4$ \\
		\hline
		\ac{BD}-\ac{RIS} size & $N_y N_z = 8 \times 8 = 64$ \\
		\hline
		Inter-element spacing $d_h = d_v$ & $\lambda/2$ \\
		\hline
		AoD, AoA (one sector of a cell) &
{\small$
\begin{aligned}
&\phi^{r}_{\mathrm{BS}},\,
\phi^{l}_{\mathrm{RIS_D}},
\phi^{r}_{\mathrm{RIS_A}},\,
\phi^{l}_{\mathrm{UE}} \\
&\sim \mathcal{U}(-60^{\circ},\,60^{\circ})
\end{aligned}$}  \\
		\hline
		EoD, EoA &
{\small$
\begin{aligned}
&\theta^{r}_{\mathrm{BS}},\,
\theta^{l}_{\mathrm{RIS_D}},
\theta^{r}_{\mathrm{RIS_A}},\,
\theta^{l}_{\mathrm{UE}} \\
&\sim \mathcal{U}(90^{\circ},\,130^{\circ})
\end{aligned}$} \\
		\hline
		Tranmit power  $P_T$ & 1 \\
		\hline
		Number of users & $1$ \\
		\hline
		Channel realizations & $1000$ \\
		\hline
		Complex path gains &  $\alpha_r, \beta_l \sim \mathcal{N}\left( 0,1\right)$ \\
		\hline
		Number of multipaths & $R = 2$, $L = 2$ \\
		\hline
	\end{tabular}
	\label{tb:simul}
\end{table}
We evaluate the proposed \ac{FORTE} and \ac{FORPE} algorithms in terms of the \ac{NMSE} of the composite channel, defined as $\textrm{NMSE}=(\hat{\bm{E}}) = (1/P)\sum^{P}_{p=1} \left\| \bm{E}^{(p)} - \hat{\bm{E}}^{(p)} \right\|^2_{\textrm{F}}/\left\| \bm{E}^{(p)}\right\|^2_{\textrm{F}}$,
where $\hat{\bm{E}}^{(p)}$ denotes the estimated composite channel in the $p$-th realization and $P$ is the total number of channel realizations. In all simulations, we set $P=1000$, as summarized in Table \ref{tb:simul}. The total transmit power is fixed to $P_T = \SI{1}{\watt}$, and the transmit \ac{SNR} is given by $\textrm{SNR} = P_T/\sigma_n^2$.
We compare the proposed methods with two benchmark estimators, namely the \ac{LS} method \cite{Hongyu_2024} and the \ac{BTKF} method \cite{Andre_Almeida_2025}, to highlight the gains achieved by explicitly exploiting the geometric structure of the underlying channels.

The simulation setup follows the parameter choices listed in Table \ref{tb:simul}. The \ac{BS} employs a \ac{UPA} with $M=M_yM_z=4\times4=16$ antennas, while the \ac{UE} uses a \ac{UPA} with $Q=Q_yQ_z=2\times2=4$ antennas. The \ac{BD-RIS} is composed of $N=N_yN_z=8\times8=64$ reflecting elements. For all arrays, the horizontal and vertical inter-element spacings are set to $d_h=d_v=\lambda/2$. We assume a single-user scenario and consider $R=2$ and $L=2$ propagation paths on the two sides of the \ac{BD-RIS}. The azimuth angles $\phi^{r}_{\mathrm{BS}}$, $\phi^{l}_{\mathrm{RIS_D}}$, $\phi^{r}_{\mathrm{RIS_A}}$, and $\phi^{l}_{\mathrm{UE}}$ are independently drawn from the uniform distribution $\mathcal{U}(-60^{\circ},60^{\circ})$, whereas the elevation angles $\theta^{r}_{\mathrm{BS}}$, $\theta^{l}_{\mathrm{RIS_D}}$, $\theta^{r}_{\mathrm{RIS_A}}$, and $\theta^{l}_{\mathrm{UE}}$ are independently drawn from $\mathcal{U}(90^{\circ},130^{\circ})$. Moreover, the complex path gains satisfy $\alpha_r,\beta_l \sim \mathcal{N}(0,1)$, consistent with the adopted geometric channel model and following the setup in~\cite{Samimi_Rappaport_2016}.

\begin{figure}[!t]
	\centering\includegraphics[scale=0.4]{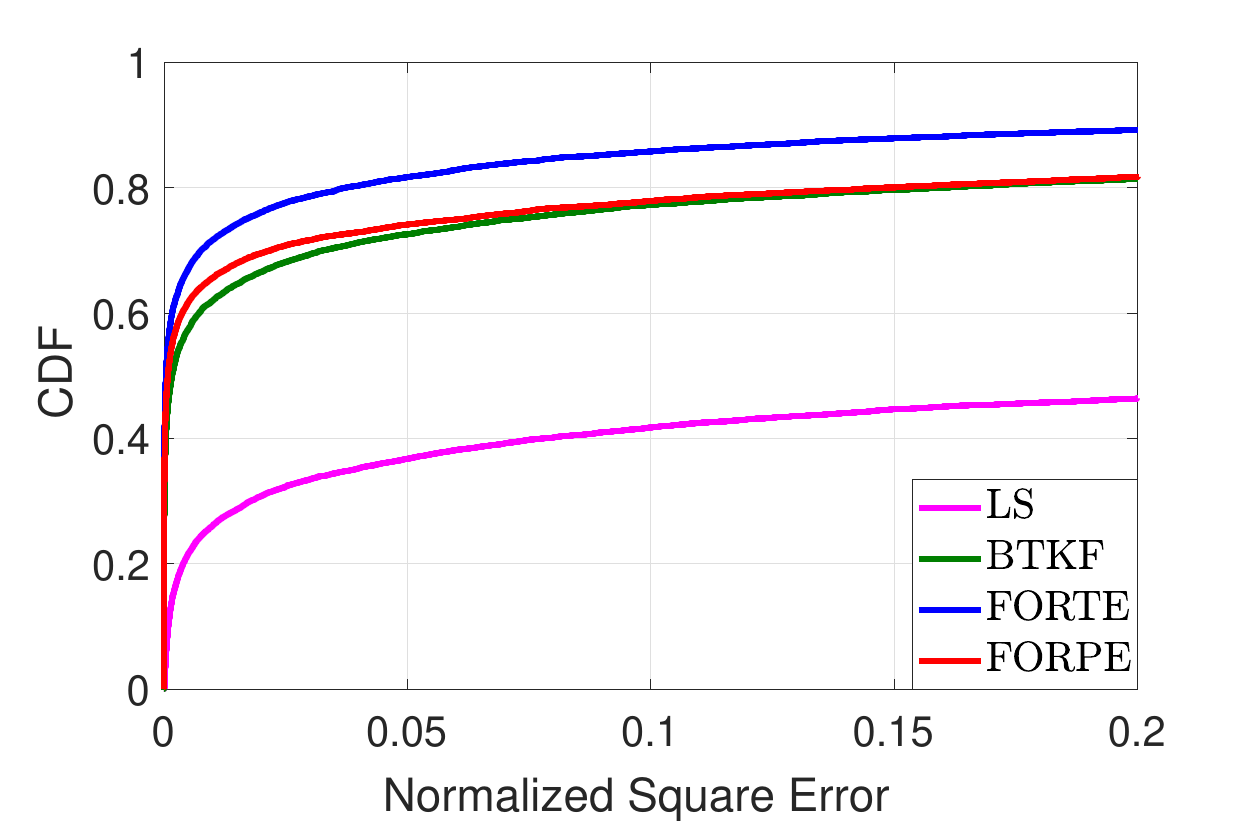}
	\caption{The \ac{CDF} of the \ac{NSE} for different methods.}
	\label{fig:CDF}
\end{figure}
Figure \ref{fig:CDF} shows the \ac{CDF} of the \ac{NSE} obtained over the entire considered \ac{SNR} range, from \SI{-30}{\decibel} to \SI{30}{\decibel}, using $1000$ channel realizations. It can be observed that the proposed \ac{FORTE} method consistently provides the best performance, outperforming the proposed \ac{FORPE} scheme as well as the benchmark \ac{BTKF} \cite{Andre_Almeida_2025} and \ac{LS} \cite{Hongyu_2024} estimators. For example, at an \ac{NSE} threshold of $0.05$, the probability that the \ac{LS} method attains this error level or lower is approximately $38\%$, whereas the \ac{BTKF}/KronF method reaches about $70\%$. Under the same operating conditions, the proposed \ac{FORPE} method attains nearly $72\%$, while the proposed \ac{FORTE} method achieves approximately $81\%$. These results confirm the superior estimation accuracy and robustness of \ac{FORTE} across the full \ac{SNR} interval considered in the simulations.

\begin{figure}[!t]
	\centering\includegraphics[scale=0.4]{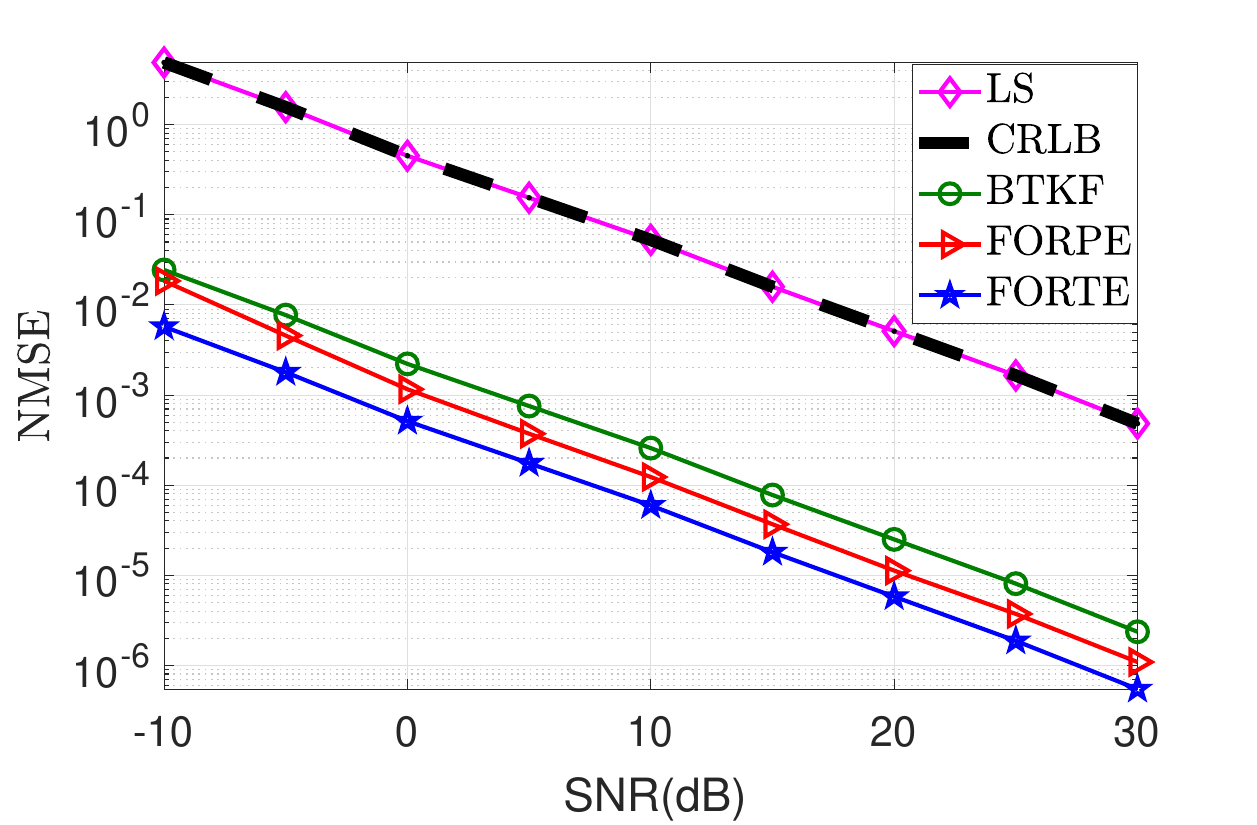}
	\vspace{-1ex}
	\caption{The \ac{NMSE} of the proposed \ac{FORTE} and the \ac{FORPE} methods with competing approaches..}
	\label{fig:NMSE}
\end{figure}

Figure \ref{fig:NMSE} compares the \ac{NMSE} performance of the proposed \ac{FORTE} and \ac{FORPE} estimators with that of the benchmark \ac{LS} \cite{Hongyu_2024} and \ac{BTKF} \cite{Andre_Almeida_2025} methods. As shown in the figure, both proposed approaches consistently outperform the reference schemes over the considered \ac{SNR} range. In particular, at \SI{5}{\decibel}, the proposed \ac{FORTE} method attains an \ac{NMSE} of approximately $10^{-4}$, while \ac{FORPE} achieves about $10^{-3.5}$. By comparison, \ac{BTKF} reaches roughly $10^{-3}$, whereas the \ac{LS} estimator remains around $10^{-1}$. This gain stems from the proposed methods explicitly exploiting the geometric structure of the underlying channel and embedding it in a tensor model, thereby enabling stronger noise mitigation during iterative estimation. In contrast, the \ac{BTKF} method leverages only the tensor structure of the composite channel, whereas the \ac{LS} method does not exploit any structural information. It is also worth noting that the \ac{LS} estimator, being tailored to unstructured composite channels, approaches the theoretical \ac{CRLB} derived for that unstructured setting.

\begin{figure}[!t]
	\centering\includegraphics[scale=0.4]{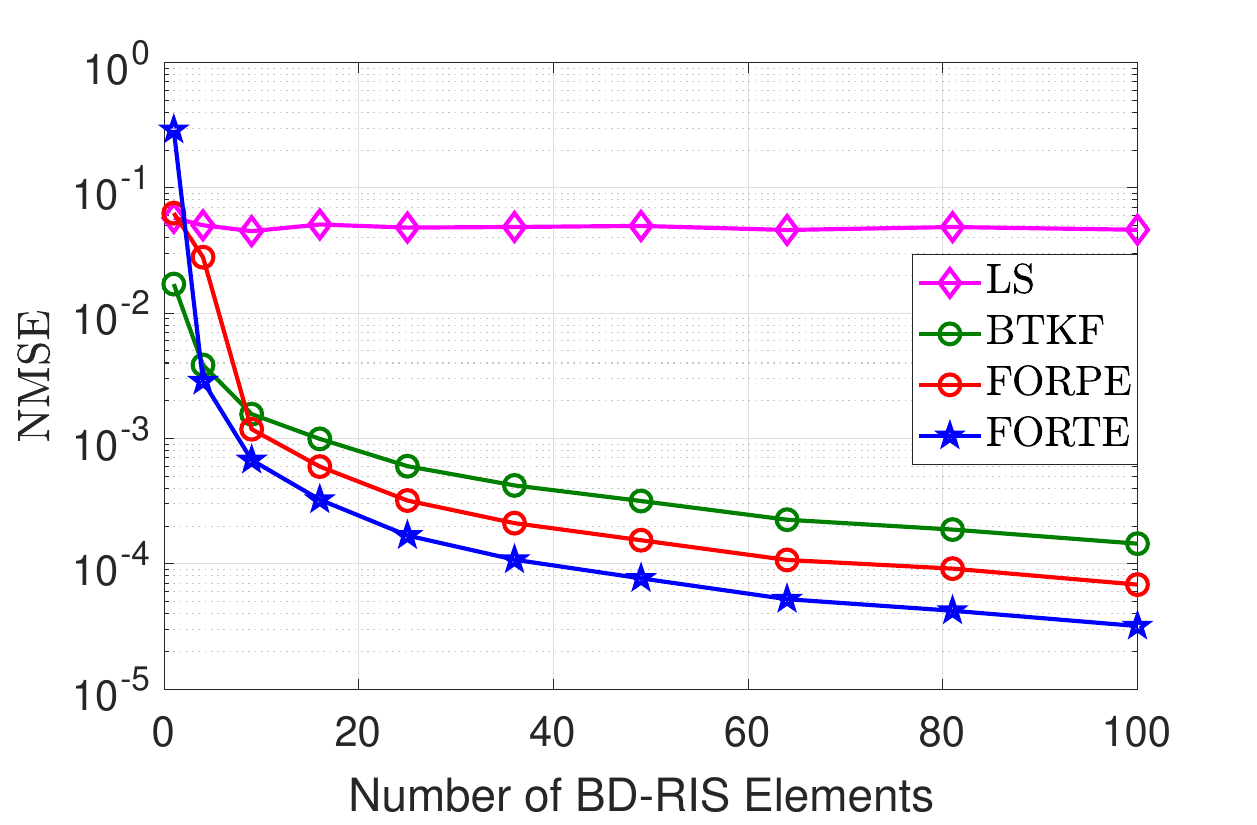}
	\vspace{-1ex}
	\caption{\ac{NMSE} based performance comparison by varying number of \ac{BD}-\ac{RIS} elements $N$ assuming \ac{SNR} \SI{10}{\decibel}..}
	\label{fig:Varying_N}
\end{figure}

Figure \ref{fig:Varying_N} illustrates the \ac{NMSE} performance as a function of the number of \ac{BD}-\ac{RIS} reflecting elements $N$ at an \ac{SNR} of \SI{10}{\decibel}, using $1000$ channel realizations. As shown in the figure, increasing the number of reflecting elements improves the performance of the proposed \ac{FORTE} and \ac{FORPE} methods, whereas the \ac{LS} estimator exhibits little to no improvement. This behavior is explained by the fact that, for the proposed methods, a larger \ac{BD}-\ac{RIS} yields a richer tensor structure and stronger noise-averaging effects, thereby enhancing the quality of the channel estimates. In contrast, the \ac{BTKF} method only exploits the tensor structure of the composite channel and therefore achieves more limited gains. For example, at $N=40$, the proposed \ac{FORTE} method attains an \ac{NMSE} of approximately $10^{-4}$, while \ac{FORPE} achieves about $10^{-3.5}$ and \ac{BTKF} reaches roughly $10^{-3.2}$. These results further confirm the benefit of explicitly incorporating the geometric channel structure into the proposed tensor-based estimation framework.

\begin{figure}[!t]
	\centering\includegraphics[scale=0.4
    ]{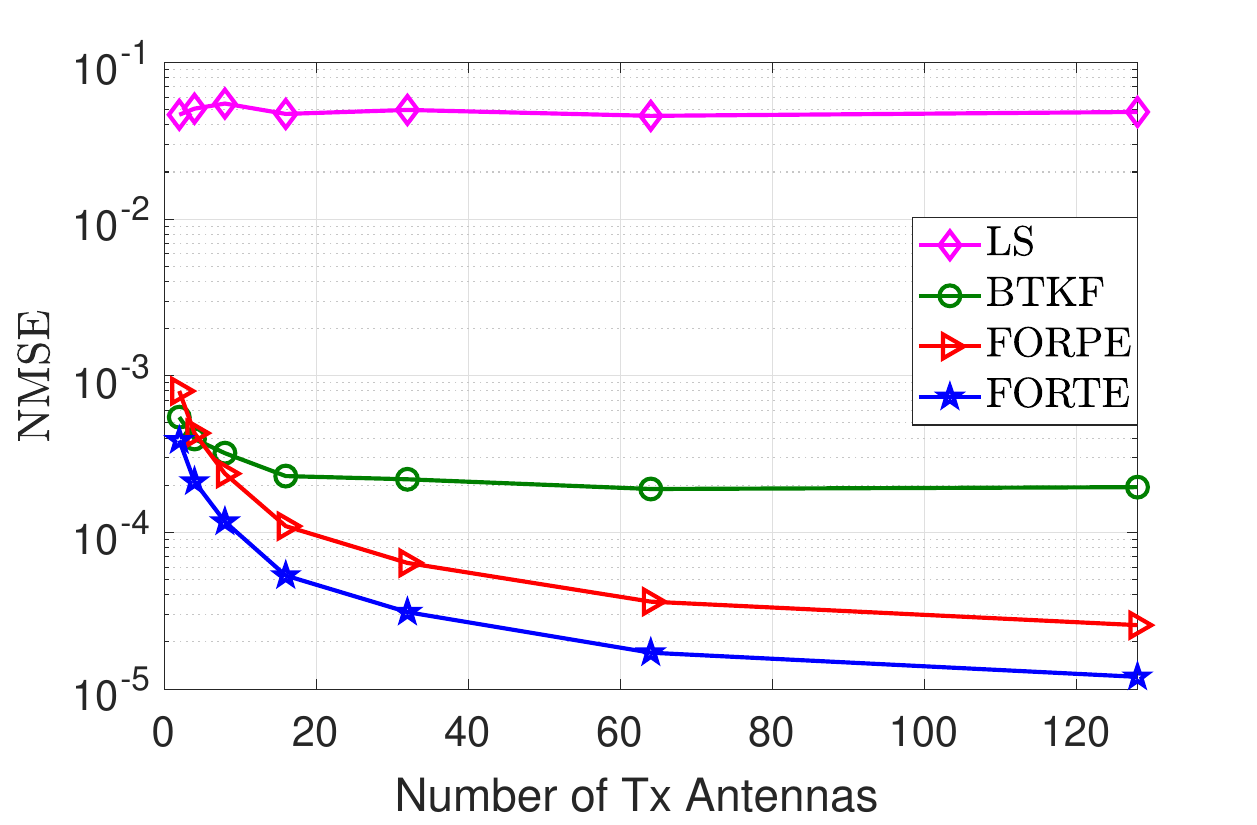}
	\vspace{-1ex}
	\caption{\ac{NMSE} based performance comparison by varying number of transmit antenna elements $M$ assuming \ac{SNR} \SI{10}{\decibel}.}
	\label{fig:Varying_M}
\end{figure}

Figure \ref{fig:Varying_M} illustrates the \ac{NMSE} performance as a function of the number of transmit antennas $M$. The values of $M$ considered are chosen as powers of two to comply with the Hadamard pilot design used in the simulations. As shown in the figure, increasing the number of transmit antennas improves the performance of both proposed methods, and \ac{FORTE} and \ac{FORPE} consistently outperform the benchmark \ac{LS} and \ac{BTKF} estimators over the entire range of $M$. This behavior can be attributed to the fact that a larger transmit array provides a richer, more structured observation model, which the proposed tensor-based methods can exploit more effectively through the geometric characterization of the channel. In contrast, the \ac{LS} method does not leverage any structural information from the composite channel, resulting in limited improvement, whereas the \ac{BTKF} method exploits only the tensor structure of the composite channel and thus achieves smaller gains than the proposed approaches.

\begin{figure}[!t]
	\centering\includegraphics[scale=0.38]{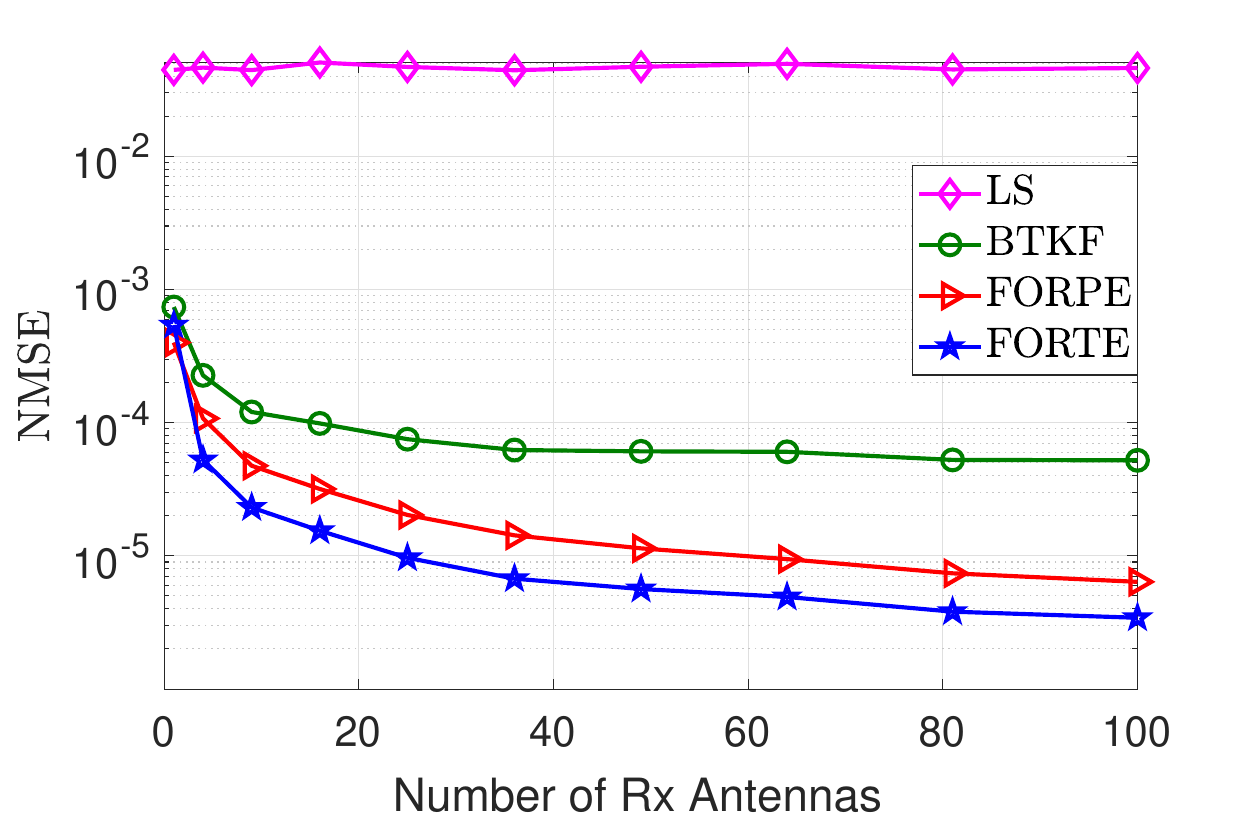}
	\vspace{-1ex}
	\caption{\ac{NMSE} based performance by varying number of receive antenna elements $Q$ assuming \ac{SNR} \SI{10}{\decibel}..}
	\label{fig:Varying_Q}
\end{figure}

Figure \ref{fig:Varying_Q} illustrates the \ac{NMSE} performance as a function of the number of receive antennas $Q$. As shown in the figure, both proposed methods, \ac{FORTE} and \ac{FORPE}, consistently outperform the classical \ac{LS} estimator and the state-of-the-art \ac{BTKF} method throughout the considered range of $Q$. This improvement is due to the proposed approaches explicitly exploiting the intrinsic geometric structure of the channel through tensor modeling, which provides an additional tensor gain and enhances estimation accuracy. In contrast, neither the \ac{LS} method nor the \ac{BTKF} method is able to benefit from this richer geometric representation to the same extent.
\begin{figure}[!t]
	\centering\includegraphics[scale=0.38]{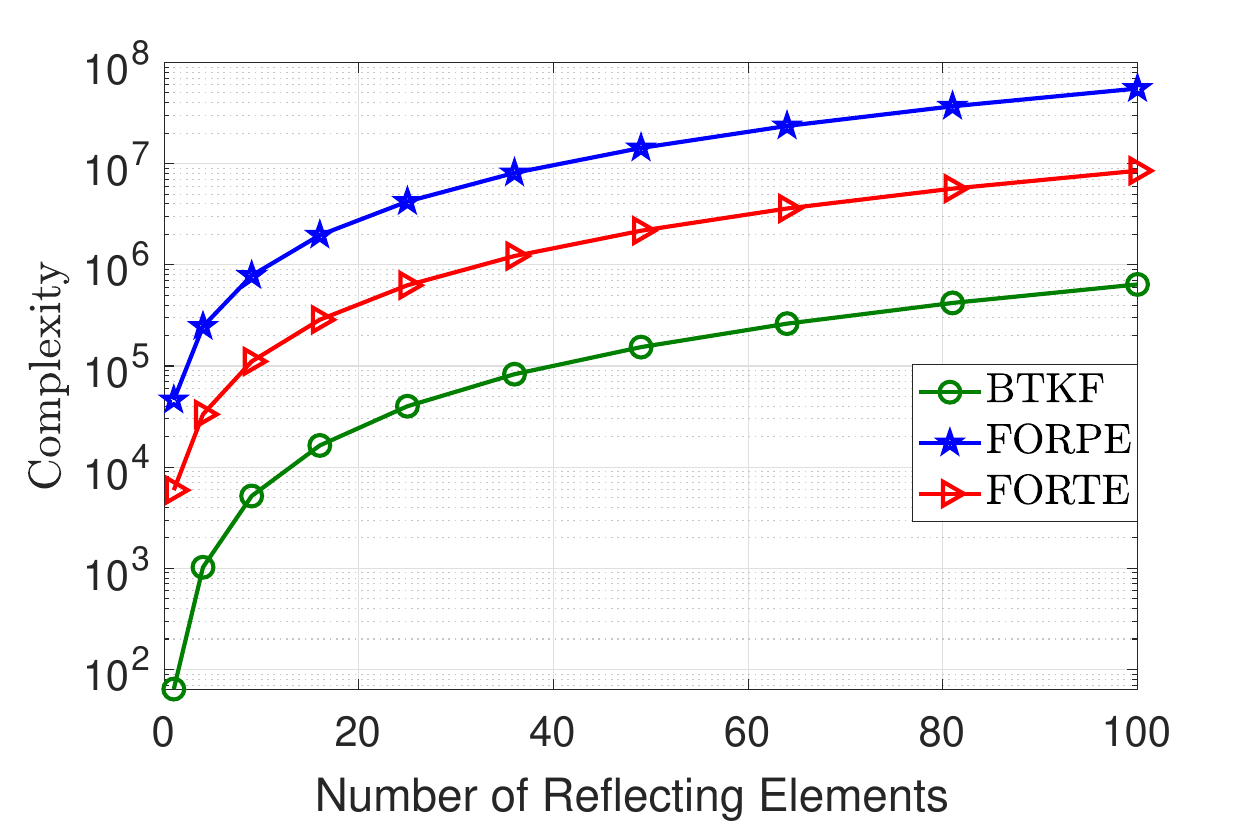}
	\vspace{-1ex}
	\caption{ Computational complexity of \ac{FORTE} and \ac{FORPE}.}
	\label{fig:complexity}
\end{figure}

Figure \ref{fig:complexity} compares the computational complexity of the proposed \ac{FORTE} and \ac{FORPE} methods with that of the benchmark \ac{LS} \cite{Hongyu_2024}  and \ac{BTKF} estimators \cite{Andre_Almeida_2025}. Specifically, the complexity curves for \ac{FORTE} and \ac{FORPE} correspond to the plain-vanilla \ac{ALS} implementations described in Algorithms \ref{algorithm_forte} and \ref{algorithm_forpe}, respectively. As illustrated in Fig.~\ref{fig:complexity}, both proposed tensor-based schemes are more computationally demanding than the closed-form \ac{BTKF} method because they rely on iterative updates and repeated least-squares subproblems. Nevertheless, this additional complexity is justified by the superior estimation performance observed in the previous numerical results. It is also worth emphasizing that all compared methods start from an initial \ac{LS} estimate, which serves as a common first-stage preprocessing step.

\begin{figure}[!t]
	\centering\includegraphics[scale=0.38]{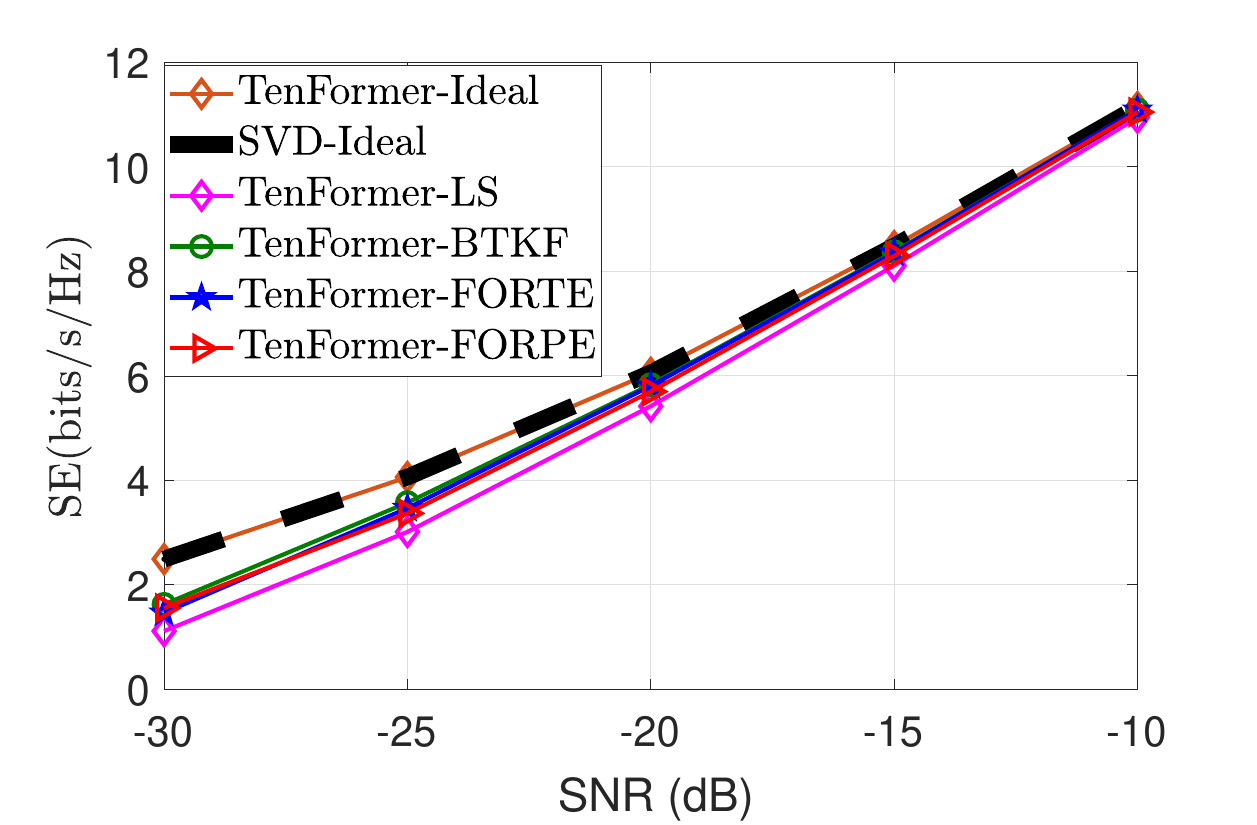}
	\vspace{-1ex}
	\caption{SE performance of \ac{TenFormer} in comparison with the benchmark \ac{SVD}-based method \cite{Emil_ozlem_2025}.}
	\label{fig:SE}
\end{figure}
\begin{figure}[!t]
	\centering\includegraphics[scale=0.38]{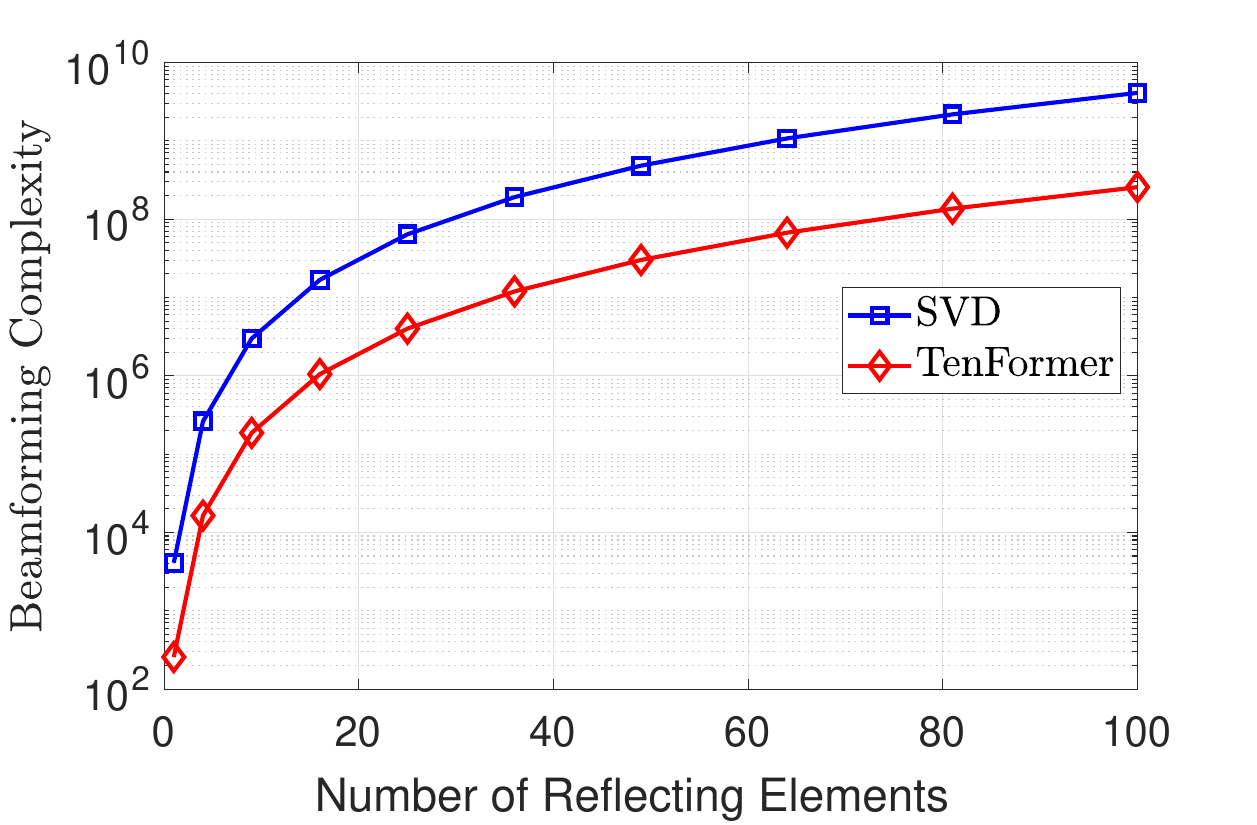}
\vspace{-1ex}
	\caption{ Beamforming complexity comparison of the proposed \ac{TenFormer} method and the competing method \cite{Emil_ozlem_2025}.}
	\label{fig:TenFormer}
\end{figure}

Figure~\ref{fig:SE} compares the \ac{SE} achieved by the proposed \ac{TenFormer} method and the benchmark scheme in \cite{Emil_ozlem_2025}, considering both true and estimated channel knowledge. As observed in the figure, the two methods exhibit similar \ac{SE} performance, since both rely on \ac{SVD}-based designs for the precoder, combiner, and scattering matrix when ideal composite channel information is available. Nevertheless, the proposed \ac{TenFormer} method offers an important practical advantage: by exploiting the intrinsic geometric structure of the channel, it enables the precoder, combiner, and scattering matrix to be obtained in parallel with lower complexity.
Figure~\ref{fig:TenFormer} compares the computational complexity of the joint design of the precoder, combiner, and scattering matrix for the proposed \ac{TenFormer} method and the \ac{SVD}-based benchmark in \cite{Emil_ozlem_2025}. As illustrated in the figure, the proposed method achieves a significantly lower computational cost by exploiting the higher-order tensor representation of the composite channel, thereby enabling the beamforming variables to be computed in parallel rather than via a more costly sequential procedure. This advantage substantially reduces the overall processing burden. For instance, when the number of \ac{BD-RIS} elements is approximately 80, the proposed \ac{TenFormer} method is about 15 times less complex than the benchmark scheme, assuming the same number of antenna elements at the \ac{BS} and the \ac{UE}.


\section{Conclusions}
\textcolor{black}{This paper investigated channel estimation and beamforming design for \ac{BD-RIS}-assisted \ac{MIMO} systems from a deconstructive tensor modeling perspective. Instead of treating the composite \ac{BD-RIS} channel as an unstructured high-dimensional matrix, we decomposed it into its directional tensor factors, thereby revealing the rich multilinear structure induced by the \ac{BS}, the \ac{UE}, and the two \ac{BD-RIS} sides.} Two tensor-based estimators were proposed: \ac{FORTE}, based on a fourth-order Tucker model, and \ac{FORPE}, based on a fourth-order PARAFAC model. 
The proposed estimators outperform competing methods across different \ac{SNR} regimes and system dimensions. In particular, \ac{FORTE} achieved the best \ac{NMSE} performance by partially exploiting the channel geometry, while \ac{FORPE} offered good accuracy with a unique estimation of channel factors. We also formulated the tensor-based \ac{TenFormer} beamforming, which achieves spectral efficiency comparable to the \ac{SVD} benchmark while incurring lower computational complexity. \textcolor{black}{Overall, deconstructing the composite \ac{BD-RIS} channel into interpretable directional factors is useful for accurate channel estimation and provides a modeling foundation for sensing-oriented parameter extraction and system optimization in future \ac{BD-RIS}-aided wireless systems.}}

\renewcommand\baselinestretch{.83}

\bibliographystyle{IEEEtran}
\bibliography{ref.bib}

\end{document}